\documentclass[aps,prd]{revtex4}
\usepackage{latexsym}

\begin{document}

\title{ From Gravitons to Gravity:  Myths and Reality}

\author{T.~Padmanabhan}
\affiliation{IUCAA, 
Post Bag 4, Ganeshkhind, Pune - 411 007\\
email: nabhan@iucaa.ernet.in}

\begin{abstract}
There is a general belief, reinforced by statements in standard textbooks, that:  (i) one can obtain
the full non-linear Einstein's theory of gravity by coupling a  massless, spin-2 field $h_{ab}$ self-consistently
to the total  energy momentum tensor, including its own; (ii) this procedure is unique and leads to Einstein-Hilbert action and (iii) it only uses standard concepts in  Lorentz invariant field theory and does not involve any geometrical assumptions.
After providing several reasons why such  beliefs are suspect --- and critically re-examining 
several previous attempts --- we provide  a detailed analysis aimed at clarifying the situation. First, we prove that it is \textit{impossible} to obtain the Einstein-Hilbert (EH) action, starting from the 
standard action for gravitons in linear theory and iterating repeatedly. This result follows from the fact that EH action has a part (viz. the surface term arising from second derivatives of the metric tensor) which is non-analytic in the coupling constant, when
expanded in terms of the graviton field. Thus, at best, one can only hope to obtain the remaining, quadratic, part of the EH Lagrangian (viz. the $\Gamma^2$ lagrangian) if no additional assumptions are made. Second, we
use the
Taylor series expansion of the action for Einstein's theory,  to identify the tensor $\mathcal{S}^{ab}$, to which the graviton field $h_{ab}$ couples to the lowest order (through a term
of the form $\mathcal{S}^{ab}h_{ab}$ in the lagrangian). We show that the second rank tensor $\mathcal{S}^{ab}$  is {\it not} the conventional energy momentum tensor $T^{ab}$ of the graviton and provide an explanation for this feature. Third, we construct
the full nonlinear Einstein's theory with the source being spin-0 field, spin-1 field or  relativistic particles by explicitly
coupling the spin-2 field to  this second rank tensor $\mathcal{S}^{ab}$ order by order and summing up the infinite series. 
Finally, we construct the theory obtained by self consistently coupling $h_{ab}$ to the 
conventional energy momentum tensor $T^{ab}$ order by order and show that this does {\it not} lead to Einstein's theory. The implications  are discussed.
 
\end{abstract}

\maketitle

\section{ Introduction and Motivation}
\subsection{Conventional wisdom ......}

 The two classical fields --- electromagnetism and gravity --- are described by a vector field
 and second rank symmetric tensor field, respectively. Considerations based on Lorentz group
 suggest interpreting them (when suitable restrictions are imposed) 
 as corresponding to massless spin-1 and spin-2 fields.
  The vector field $A_i$ couples to a conserved current
 $J_i$ but does not contribute to this current. (That is, the photon does not carry charge.)
 In contrast, the tensor field is believed to be coupled to the energy momentum tensor;  since the field itself
 carries energy, it has to couple to itself in a non linear fashion. (The situation is similar to
 Yang-Mills fields which carry isotopic charge and hence are non linear.) It may, therefore,
 be possible to obtain a correct theory for gravity by starting with a massless spin-2 field
 $h_{ab}$ coupled to the energy momentum tensor $T_{ab}$ of other matter sources to the lowest 
 order,  introducing self-coupling of $h_{ab}$ to its own energy momentum tensor at the
 next order 
 and iterating the process.  
 This will lead to a completely field theoretic description of gravity in a Minkowski background
 and is conceptually quite attractive.
 
 This attempt has a long history. The field equation for a free massless spin-2 field was originally
 obtained by Fierz and Pauli \cite{fierz}.
  The first attempt to study the consequences of coupling 
 this field to its own energy momentum tensor seems to have been by Kraichnan in unpublished
 work done in 1946-47. The first published attempt to derive the non linear coupling is by
 Gupta \cite{gupta}
  and Kraichnan published some of his results soon after \cite{kraich55}.
   Feynman
 has provided  a derivation \cite{feynman} in his Caltech lectures on gravitation during 1962-63. The problem was
 re-addressed by a clever technique by Deser \cite{deser}. 
 (This problem and related ideas have been explored from several other points of view in literature; see e.g., 
\cite{alternative}. We shall not discuss these approaches.) Virtually all these approaches claim to obtain  not only Einstein's field equations 
 but also the Einstein-Hilbert action.
 
 \subsection{ ...... And why it is suspect}
 
 This result is widely quoted in literature (see, e.g., page 424 of \cite{mtw}) and, at first sight, seems eminently reasonable. However, deeper
 examination raises several disturbing questions, if the above result is really valid.
  
 \begin{itemize}
 
 \item  In the conventional derivations, the final metric arises as $g_{ab}=\eta_{ab}+\lambda h_{ab}$
where $\lambda\propto \sqrt{G}$ has the dimension of length and $h_{ab}$ has the correct dimension of
(length)$^{-1}$ in natural units with $\hbar=c=1$. The iteration is in powers of $\lambda$, starting from the zeroth order lagrangian $L_0\simeq (\partial h)^2$ for 
spin-2 field,  which has the dimension of (length)$^{-4}$. (We have dropped the tensor indices to simplify the notation.) The final result in all the published works is the Einstein-Hilbert lagrangian $L_{EH}=(1/4\lambda^2)R$. Since the scalar curvature has the structure $R\simeq (\partial g)^2+\partial^2g$, substitution of $g_{ab}=\eta_{ab}+\lambda h_{ab}$ gives to the lowest order:
\begin{equation}
L_{EH}\propto \frac{1}{\lambda^2}R\simeq (\partial h)^2+\frac{1}{\lambda}\partial^2h
\end{equation}
Thus the full Einstein-Hilbert lagrangian is non-analytic in $\lambda$! {\it It will be quite surprising if, starting from $(\partial h)^2$ and doing a honest iteration on $\lambda$, one can obtain a piece which is non-analytic in $\lambda$.} At best, one can hope to get the quadratic part of $L_{EH}$ which gives rise to the $\Gamma^2$ action but not the four-divergence term involving $\partial^2g$.

 \item 
 To carry out this programme,
 one need to identify the energy momentum tensor $T_{ab}^G$ for the graviton field $h_{ab}$ order by order in the coupling constant. At this stage, we are working in flat spacetime, Cartesian coordinates [with metric $\eta_{ab}=dia (-1,1,1,1)$]
with Lorentz group as the invariance group. If we are honest (and do not use anything we learnt in our
general relativity course!), we must provide a prescription to find $T_{ab}^G$ within this context. There is indeed
a natural conserved second rank tensor which arises from Lorentz symmetry, usually called the canonical energy momentum tensor. This tensor, unfortunately, is not symmetric. It can be made symmetric but the procedure is not unique. For every choice of $T_{ab}^G$ one can one obtain a nonlinear theory clearly showing that further choices are to be made somewhere along the line.

\item A sharper way of stating the above difficulty is the following: The same textbooks which assert that
Einstein's theory can be obtained by coupling $h_{ab}$ to itself self consistently will also state in some other section (in ref. \cite{mtw}, this happens in page 467)  that gravitational field does not have a well defined energy momentum tensor!. It will be rather strange if a  unique energy momentum tensor exists for gravitational field order by order in the
perturbation series but somehow ``disappears" when all the terms are summed up. 
(The non-uniqueness of energy momentum tensor for Einstein's theory is well known and is extensively discussed in literature; see e.g., \cite{gravtik}.)

\item In implementing this program, one needs to be clear whether general covariance is an \textit{assumption}
or a \textit{result}. The starting point --- Lorentz invariant field theory in flat spacetime with metric $\eta_{ab}$ ---
has no notion of general covariance. If the source of the final equations is an energy momentum tensor which is {\it assumed} to be generally covariant, it is equivalent to assuming that the left hand side of the equations
is generally covariant. It is then no big deal to obtain Einstein's theory, if we are prepared to {\it assume}
general covariance. [It is sometimes claimed that the gauge invariance of spin-2 field 
under $h_{ab}(x) \to h_{ab}(x) + \partial_a \xi_b (x) + \partial_b \xi_a (x)$, ``becomes" the general covariance of the full theory. This is simply wrong; see the discussion around Eq.~(\ref{fourdiv}) in  Section III below.]

\item A term in the lagrangian proportional to $\lambda h_{ab}T^{ab}$ where $T^{ab}$ is due to {\it external} matter fields (assumed to be independent of $h_{ab}$ to this order), will lead to the equation of motion of the type $\partial^2 h=\lambda T$. Hence a coupling of the type
 $\lambda h_{ab}T^{ab}$ is equivalent to requiring the source being $T_{ab}$. Consider now the coupling of gravity to itself through a term of the type $\lambda h_{ab}\mathcal{S}^{ab}(h)$ where $\mathcal{S}^{ab}$ explicitly depends on the graviton field $h_{ab}$. When this term is varied with respect to $h_{ab}$ to get the equations of motion we will obtain {\it two} terms:
$\mathcal{S}^{ab}+(\partial \mathcal{S}^{ij}/\partial h_{ab})h_{ij}$ both of which will act as a source to gravity at next order. If we want the {\it source} to be the energy momentum tensor of graviton field, $T_{ab}$, then the {\it coupling}  cannot be of the form
$h_{ab}T^{ab}(h)$ since this will lead to the wrong source. Thus we need to find out the form of the tensor 
$\mathcal{S}^{ab}$ --- a question which does not seem to have attracted any attention in the literature.
(We will see that $\mathcal{S}^{ab}$ is an interesting object in its own right.)

\end{itemize}
 
  None of the previous derivations addresses these issues and most of them downplays the role of {\it assuming}
  general covariance. 
 All these attempts make different tacit assumptions and it is difficult to judge which of these
 derivations can be thought of as ``from first principles'' in the sense that it is completely independent
 of our knowledge of the end result. This difficulty becomes apparent when one follows
  the details of many of these derivations. The technology used is very strongly influenced by the 
known final result.  For example, Kraichnan's pioneering work explicitly uses a term like 
 $\eta_{ab} R^{ab} (\eta)$ 
 (where $\eta_{ab}$ is the Minkowski metric and $R^{ab}$ is the Ricci tensor)
 cleverly to obtain the result,
 [see equations 13-17 of \cite{kraich55}]
  in spite of the fact that $ R^{ab} (\eta)$ vanishes for the flat
 metric $\eta_{ab}$ ! It is impossible (at least for the author) to imagine that someone could have
 ``guessed'' this form for the action without knowing the result. Feynman's derivation also suffers from 
 several shortcomings. To begin with, it is considerably less general than the one by Kraichnan
 since Feynman assumes a particular form for the matter action and a coupling. But more relevant
 to our discussion is the manner in which he constructs the solutions to a consistency condition
 (see the discussion in section 6.3 and 6.4 of \cite{feynman}).
  Since this approach 
  \emph{assumes} general covariance ( in the form $\nabla_a T^{ab}=0$) and
   relies heavily on constructing generally covariant scalars, it is 
 predestined to give Einstein-Hilbert action. The by far cleverest mathematical procedure was the 
 one employed by Deser, in which he exploits the fact that, with a suitable choice of variables, the 
 gravitational action becomes a cubic polynomial allowing the iteration to stop at a finite
 order. To achieve this mathematical economy, he  has to start with the Palatini variational form
 (see his equation 2) based on the Lagrangian $f^{ab} R_{ab}$ where $f^{ab} = \sqrt{-g} g^{ab}$
 is the preferred variable. It is no surprise that he obtains $\sqrt{-g} R$ as the final result.
 (This is the {\it only} previous work that actually attempts an iteration; we shall comment on Deser's derivation in more detail in Section VI C.)
 
 In particular, \textit{the fact that the action for the final theory contains
  the second derivatives of the field is  always put in by hand}. Kraichnan's work has 
 this explicitly; Feynman's derivation assumes a   condition equivalent to general covariance
 to obtain solutions to a functional constraint and he  explicitly chooses the ``simplest" generally
 covariant scalar thereby getting
  $\sqrt{-g}R$;  Deser starts with an expression which is numerically
 same as $\sqrt{-g}R$ but since he uses the first order form of the variation, the question of \emph{second}
 derivatives is not directly applicable until, of course, when the final result is obtained. Thus, we really
 do not know whether the original program 
 (of coupling the field to its own energy momentum tensor
 order by order and iterating the result) can be made to yield any other form of action principle 
 even when the field equations are the same.  
 
 One may be tempted to argue that most of the issues and objections raised above are irrelevant in the strictly classical context. In classical general relativity, one could argue, what matters is the equations of motion and not action functional. This, however, is a rather restricted point of view and one needs to realise that the true world is quantum mechanical and if one can gain insight into the nature of quantum theory from the structure of classical action functional, it is worth exploring. Of course,  
 quantum
 theory has taught us that action functionals are as important (if not more) as the field equations. In the case of gravity, there are two action functionals which are of primary
 relevance. The first is the Einstein-Hilbert action which uses the Lagrangian $R\sqrt{-g}$ and 
 the second is the $\Gamma^2$ action involving only the squares of the first derivatives. It can be
 shown 
 that these two actions can be thought of as providing the momentum
 representation and the coordinate representation (respectively) of the theory and differ by a surface term which is directly related to the entropy of horizons in the semi classical theory and has been the basis of a series of investigations \cite{tppapers}.
 Since the existence of horizons is probably the most remarkable feature of classical gravity that could
 serve as a link with the quantum description of spacetime,
  it is important to try and understand whether this surface term can arise from the spin-2 field approach.

\subsection{Plan of the paper}

We will try to address these issues in as straight forward (``dumb") a manner as possible. Section II reviews  the background material related to spin-2 field and few important results, needed later, are obtained.  In particular,
we introduce a new second rank tensor $\mathcal{S}^{ab}$ associated with  any matter lagrangian
which can be obtained by a well-defined procedure.
This tensor, in general, is  different from the standard energy momentum tensor 
$T^{ab}$
 but coincides with the energy momentum tensor for relativistic particle, spin-0 field or spin-1 field. 

In Section III, we start with an action for Einstein gravity for the metric $g_{ab}=\eta_{ab}+\lambda h_{ab}$,
expand it is in a functional Taylor series
in $h_{ab}$ and determine the form of the self-coupling at the lowest order. 
We show that, to the lowest non trivial order, the coupling is of the form $h^{ab} \mathcal{S}_{ab}$ where
$\mathcal{S}_{ab}$ is the quantity introduced in Section II. We also exhibit the non-analytic nature of the
Einstein-Hilbert lagrangian in $\lambda$ and prove that Einstein-Hilbert lagrangian can never be obtained by an iteration in $\lambda$.

In Section IV we provide a general procedure for coupling the field $h_{ab}$ self consistently
 to \emph{any} second rank tensor  which can be expressed as functional derivative of matter action. This leads to a well defined ``rule" for coupling the field $h_{ab}$ to matter fields.
We first use it with the tensor $\mathcal{S}^{ab}$ we have defined in Section II  and show that it leads to a generally covariant lagrangian for relativistic particle, spin-0 field or spin-1 field  but not in general.  We then use the same prescription to couple $h_{ab}$ to itself and show that
{\it the resulting theory is Einstein's theory} (Section IV A). Thus, at least in the limited case of spin-2 field interacting with
relativistic particle, spin-0 field or spin-1 field [the only cases in which we have any observational evidence for gravitational theory!], we have a iterative procedure for obtaining the full theory when the self-coupling of $h_{ab}$ is {\it not} to the energy momentum tensor. In Section IV B we repeat the analysis by coupling 
$h_{ab}$ using the standard
energy momentum tensor $T^{ab}$. In the case of {\it all } matter fields, this leads to the standard generally covariant action. But when we use this prescription for coupling $h_{ab}$ to itself, we do {\it not} get Einstein's theory but a more complicated one which explicitly depends on the background Lorentzian metric 
or the field $h_{ab}$. 
We shall also show (in Section IV C) that previous results, when properly analysed, agree with our claims.

The analysis in this paper goes contrary to the conventional wisdom and Section V discusses the issues which arise from this work.

 \section {Action and Energy momentum tensor for the spin-2 field}
 
In this section we will  collect together the results which are required later.  (A more
pedagogical description is provided in Appendix A; this may be useful since  the results are somewhat scattered in the literature \cite{spintwo}). 
The action for the non-interacting, massless, spin-2 field $h_{ab}$ is built out of scalars which are quadratic in
the derivatives $ \partial_a h_{bc}$. The most general expression will be the sum of different scalars obtained by
contracting  pairs of indices in $ \partial_a h_{bc}\partial_i h_{jk} $ in different manner. If we assume that
the field equations should be invariant under the gauge transformation:
\begin{equation}
 h_{ab}(x) \to h_{ab}(x) + \partial_a \xi_b (x) + \partial_b \xi_a (x).
 \label{gt}
 \end{equation}
then the
 resulting expression for the quadratic part of the action can be written in different, equivalent, forms:
\begin{eqnarray}
A_h &=&\frac{1}{4} \int d^4x\,  \partial_a h_{bc}\partial_i h_{jk} 
\left[ \eta^{ai}\eta^{bc}\eta^{jk}
 -\eta^{ai}\eta^{bj}\eta^{ck}
+2\eta^{ak} \eta^{bj}\eta^{ci}
 -2\eta^{ak}\eta^{bc}\eta^{ij}\right]\nonumber\\
 &=& \frac{1}{4} \int d^4x\, \left[ \partial_i h^a_a \partial^i h^j_j 
 - \partial_a h_{bc} \partial^a h^{bc} 
 + 2 \partial_a h_{bc} \partial^c h^{ba}
 - 2 \partial_a h^b_b \partial_i h^{ia}
 \right]\nonumber\\
 &=& \frac{1}{4} \int d^4x\, \left[ \frac{1}{2}\partial_i \bar h^a_a \partial^i \bar h^j_j 
 - \partial_a \bar h_{bc} \partial^a \bar h^{bc} 
 + 2 \partial_a \bar h_{bc} \partial^c \bar h^{ba}
 \right]; \qquad \bar h_{ab}\equiv h_{ab}-\frac{1}{2}\eta_{ab}h^i_i
\label{spintwoact}
\end{eqnarray}
[If we assume that the action is quadratic in the first derivatives and gauge invariant, its form is uniquely
given by the above equation, except for one {\it very specific} four divergence term which can be added.  This is discussed in Appendix A; see Eq. (\ref{divamb}).]
We shall use the more compact notation:
\begin{equation}
A_h =\frac{1}{4} \int d^4x\,  \partial_a h_{bc}\partial_i h_{jk} M^{abcijk}(\eta^{mn})
\label{spintwo}
\end{equation}
where the tensor $M^{abcijk}(\eta^{mn})$  is symmetric in $bc, jk$ and under the triple exchange
$(a,b,c) \leftrightarrow (i,j,k)$ and is given by:
\begin{equation}
M^{abcijk}(\eta^{mn}) = \left[\eta^{ai} \eta^{bc}\eta^{jk}
 -\eta^{ai}\eta^{bj}\eta^{ck}
+2 \eta^{ak}\eta^{bj}\eta^{ci}
 -2\eta^{ak}\eta^{bc}\eta^{ij}\right]_{\rm symm}
 \label{formofm}
\end{equation}
where the subscript ``symm'' indicates that the expression inside the square bracket should be
suitably symmetrized in $bc, jk$ and under the exchange $(a,b,c) \leftrightarrow (i,j,k)$.
In the expression for the action, since $M^{abcijk}$ is multiplied by  $\partial_a h_{bc}\partial_i h_{jk}$,
we need not worry about symmetrization and use the expression given inside the square bracket
in Eq.~(\ref{formofm}) as it is. 
The gauge invariance of the action leads to the {\it identity}
\begin{equation}
M^{abcijk} \partial_b\partial_a\partial_i h_{jk}=0.
\label{cond}
\end{equation}

 To the lowest order, we can couple $h_{ab}$ to other fields by adding an interaction lagrangian of
 the form $(\lambda/2)T^{ab}h_{ab}$ where $T^{ab}$ is some tensor built out of the matter variables and
$\lambda$ is a coupling constant.  The total action will be:
 \begin{equation}
A_{tot} =\frac{1}{4} \int d^4x\,  \partial_a h_{bc}\partial_i h_{jk} M^{abcijk}(\eta^{mn})+
\frac{\lambda}{2} \int d^4x\,  h^{ab}T_{ab}+A_{matter}
\label{htint}
\end{equation}
Obviously, only the symmetric part of $T^{ab}$ is relevant for this coupling.
 The variation of $h_{ab}$ will now lead to the field equation $M^{abcijk} \partial_a\partial_i h_{jk} = \lambda T^{bc} $,  The condition in Eq.(\ref{cond}) now implies that
 $\partial_a T^{ab}=0$. Thus the field described by the action in
Eq.~(\ref{spintwoact}) can only be sourced by a conserved, symmetric part of a second rank tensor. 
The above fact --- which is, of course, fairly standard --- shows the intimate connection
 between the conservation of the source and the gauge invariance of the field. 
 
 It is this conservation law $\partial_a T^{ab}=0$ which leads to an inconsistency if we assume that $T^{ab}$ is
 the standard expression for energy momentum tensor for matter fields. 
 When the {\it matter variables} are varied, the equation of motion will now be affected by $h_{ab}$
 because of the $(\lambda/2)T^{ab}h_{ab}$ coupling. But the condition $\partial_a T^{ab}=0$ is equivalent to
 the equations of motion for matter field, when it is  {\it unaffected} by $h_{ab}$. Hence, in general, it will not be possible to satisfy
 both these conditions and find consistent set of solutions. The conventional wisdom is to attempt to find a consistent theory in which
 the field equations for $h_{ab}$ should imply not the condition $\partial_a T^{ab}=0$ but a modified one of
 the form $\partial_a (T^{ab}+t^{ab})=0$ where $t_{ab}$ is the energy momentum tensor for the spin-2 field.
 This will require coupling the field to its own energy momentum tensor recursively and the hope is to
 show that --- when the recursion is carried out to infinite orders --- the resulting theory will be Einstein's gravity.  This brings us to the question of defining the energy momentum tensor for the spin-2 field.
  
  For any system described by a Lorentz invariant Lagrangian $L(\phi_A, \partial_a \phi_A)$
  where $\phi_A$  denotes a generic matter field with $A$ representing possible tensor indices,
  one can show that 
  \begin{equation}
  \partial_b\left[ \partial_a \phi_A \left(\frac{\partial L}{\partial (\partial_b \phi_A)} \right) - \delta^b_a L \right]
  =0
  \end{equation}
  when the equation of motion is satisfied  \cite{belinfante}.  This allows us to define  
   an infinite number of conserved second rank tensors  
  of the form 
  \begin{equation}
  T^{ba} \equiv \left[ \partial^a \phi_A \left(\frac{\partial L}{\partial (\partial_b \phi_A)} \right) - \eta^{ab} L
   \right] + \partial_c \psi^{cba}
   \end{equation}
   where $\psi^{cba} = - \psi^{bca}$ is an arbitrary third rank tensor anti symmetric in the first two 
   indices, so that $\partial_c \partial_b \psi^{cba} =0$ identically. It is possible to choose
   $\psi^{cba}$ in infinite number of ways and still ensure that $T^{ba}$ is symmetric. Thus Lorentz invariant
   field theories possess infinite number of conserved symmetric second rank tensors any of which can be
   legitimately thought of  as an energy momentum tensor. For the spin-2 field this prescription gives
   \begin{equation}
   T^{pq}  =\frac{1}{2} M^{pbcijk} \partial_ih_{jk}  \partial^q h_{bc} - \eta^{pq} L + \partial_c \psi^{cpq}
   \label{tpqdef}
   \end{equation}
 This non-uniqueness shows that it is not possible to proceed further without making extra assumptions
 regarding the form of $T^{pq}$. 
 
One needs to be clear about the {\it different} kinds of ambiguity in the definition of $T^{pq}$. The first ambiguity is
in the choice of  $\psi^{cpq}$. The second ambiguity has to do with the fact that we can add to our Lagrangian 
a total divergence with 
 a undetermined coefficient [as shown in Eq.(\ref{divamb}) of Appendix A]. This changes the form of $T^{pq}$.  Third, the $T^{pq}$ defined in Eq.~(\ref{tpqdef}) is \emph{not}
 gauge invariant.
 In fact, one can prove a general theorem \cite{gtoftpq} that the energy momentum tensor for the 
 spin-2 field cannot be made gauge invariant for any choice of $\psi^{cpq}$.
 This raises serious questions about whether the resulting theory after infinite iteration
 will possess any trace of the original gauge symmetry. 
 If one were to be honest, in the sense that {\it no structures other than those sanctioned by Lorentz
 invariant field theory are to be used}, then it is not possible to proceed  any further and obtain a {\it unique} 
nonlinear theory.

 Let us, however,  reduce the standards of honesty and introduce another definition of a second rank
  symmetric tensor (which is based on what we learnt in our general relativity course but we won't mention it!) along the 
  following lines: 
We note that the Lagrangian for any field also depends
  on the Lorentz metric $\eta^{ab}$, i.e, $ L=L(\phi_A, \partial_a \phi_A,\eta_{ab})$. The functional derivative
  of the action $A_{matter}$ with respect to $\eta_{ab}$ will define a symmetric, second rank tensor which we can attempt to use. But since $\eta_{ab}={\rm dia}\ (-1,1,1,1)$ is a
constant $(\delta A/\delta \eta_{ab})$ is mathematically ill defined and the functional derivative actually depends on the rule for its definition.  We shall see below  that several rules are possible but let us first consider the conventional wisdom again. 

We begin by
 noting that, even though we are in flat spacetime,  we can use any set of coordinates to describe the physics. Let us assume that, in a curvilinear coordinate system we choose,  the spacetime metric 
 is $\gamma_{ab}(x)$.  We will further  assume that the action in the curvilinear coordinates is obtained by replacing $\eta_{ab}$ by $\gamma_{ab}$, ordinary derivatives into covariant derivatives and  changing the volume element from $d^4x$ to
$d^4x \sqrt{-\gamma}$. Thus the action
 has a kinematic dependence on $\gamma_{ab}$ which we shall explicitly exhibit by writing it as 
 \begin{equation}
 A(\phi_A,\partial\phi_A,\eta_{ab})\to
 A_\nabla(\phi_A,\nabla\phi_A,\gamma_{ab})=\int d^4x \sqrt{-\gamma}\, L_\nabla(\phi_A,\nabla\phi_A,\gamma_{ab})
 \label{gencovact}
 \end{equation}
(The subscript $\nabla$ in $A_\nabla$ is to remind ourselves that, in obtaining this action, ordinary derivatives have been changed to covariant derivatives; this will turn out to be important later on.)
 It is now possible to obtain a second rank symmetric tensor $T^{ab}$ by taking the functional derivative of the action  with respect to $\gamma_{ab}$ and then setting $\gamma_{ab}=\eta_{ab}$:
\begin{equation}
\delta A_\nabla =  \frac{1}{2} \int d^4 x\, \sqrt{-\gamma} \, T^{ab} \delta \gamma_{ab} ; 
\quad
T^{ab}(x)\equiv \left[\frac{2}{\sqrt{-\gamma}}
\frac{\delta A_\nabla}{\delta \gamma_{ab} (x)}\right]_{\gamma=\eta}
\label{deftab1}
\end{equation}
More explicitly, this leads to the energy momentum tensor:
\begin{equation}
T^{ab}(x)
=\left[\frac{2}{\sqrt{-\gamma}}  \left\{\frac{\partial L\sqrt{-\gamma}}{\partial \gamma_{ab} } - 
\partial_c\left( 
\frac{\partial L\sqrt{-\gamma}}{\partial (\partial_c\gamma_{ab})} \right)\right\} \right]_{\gamma=\eta}
\label{deftab}
\end{equation}
The procedure described above provides one possible prescription for obtaining $T^{ab}$ 
in flat spacetime. Note that $\gamma_{ab}$ for us is purely a bookkeeping device and, in the end of the calculations, we shall set $\gamma_{ab}=\eta_{ab}$. 

It is rather surprising that 
this definition of $T^{ab}$ is routinely used in field theoretic approaches to gravity (like, for example, in \cite{feynman}) as though it has nothing
to do with curved spacetime. {\it This attitude is incorrect}.  The variation of $\gamma_{ab}$ to $\gamma_{ab}
+ \delta \gamma_{ab}$ for arbitrary choices of $\delta \gamma_{ab}$ takes one from flat spacetime
in curvilinear coordinates to genuine curved spacetimes.  (Just varying the coordinates in flat spacetime will only have 4 function degrees of freedom while we need 10). The evaluation of the functional derivative
in Eq.~(\ref{deftab}) requires the \emph{strong} assumption that the action in Eq.~(\ref{gencovact}) is valid 
in arbitrary curved spacetime with metric $\gamma_{ab}$. We have come a long way from the basic
concepts of Lorentz invariant quantum field theory in flat spacetime.

To use this definition for the spin-2 field, we first write the action in Eq.~(\ref{spintwoact}) in arbitrary curvilinear coordinates with metric $\gamma_{ab}$, using our rule in Eq. (\ref{gencovact}) as:
\begin{eqnarray}
A_\nabla &=&\frac{1}{4} \int d^4x\, \sqrt{-\gamma}\,  \nabla_a h_{bc}\nabla_i h_{jk} \, M^{abcijk}(\gamma^{mn})\nonumber\\
&=& \frac{1}{4} \int d^4x\, \sqrt{-\gamma}\,  \nabla_a h_{bc}\nabla_i h_{jk} 
\left[ \gamma^{ai}\gamma^{bc}\gamma^{jk}
 -\gamma^{ai}\gamma^{bj}\gamma^{ck}
+2 \gamma^{ak}\gamma^{bj}\gamma^{ci}
 -2\gamma^{ak}\gamma^{bc}\gamma^{ij}\right]
\label{habincst}
\end{eqnarray}
In this expression, the covariant derivative operator $\nabla$ is defined with respect to the metric 
$\gamma_{ab}$ and involves the first derivatives $\partial_a \gamma_{bc}$. Varying this action
with respect to $\gamma_{ab}$ will lead to a symmetric energy momentum tensor for the spin-2 field when
we use the prescription in Eq.~(\ref{deftab}).
The actual expression for this tensor is fairly complicated but 
--- fortunately --- we do not need it. We, however, stress the following fact:
Since 
$[\partial L/ \partial (\partial_c\gamma_{ab})] $ involves first derivatives of $h_{ab}$, the tensor
$T^{ab}$ will involve \emph{second derivatives} of $h_{ab}$. It can again be shown by detailed algebra
that this $T^{ab}$ is indeed of the form in Eq.~(\ref{tpqdef}) for a specific choice of $\psi^{cpq}$;
thus, our rule chooses one out of many choices in Eq.~(\ref{tpqdef}).  We can now attempt to 
obtain the non linear theory by coupling $h_{ab}$ iteratively to the energy momentum tensor defined
by Eq.~(\ref{deftab}). We shall show in Section  IV B that {\it contrary to popular belief, the resulting theory
is not Einstein's theory. }

There is, however, a  more important issue which needs to be raised as regards the procedure used to obtain Eq.~(\ref{deftab}). What we have done is essentially to introduce the curvilinear metric
$\gamma_{ab}$ into the matter action (which was originally defined in flat spacetime Cartesian coordinates)
  \emph{by a particular rule} and then evaluate the functional derivative in Eq.~(\ref{deftab}). At the 
  end of the calculation, we set $\gamma_{ab} \to \eta_{ab}$. 
The rule in Eq.(\ref{gencovact}) is strongly motivated by general covariance and, of course, leads to a generally covariant matter action in the curvilinear coordinates.
But since we do {\it not} have the right to 
  assume general covariance (and only Lorentz invariance), the rule we have specified is only
  one among many possible ways of introducing $\gamma_{ab}$ into the matter action. 

To bring this
  point sharply into focus and to derive some important consequences in the coming Sections,
  we shall introduce another rule which leads to the definition of  \emph{another} symmetric second rank
  tensor. 
  To do this, we will construct a modified  action in the curvilinear coordinates  by replacing $\eta_{ab}$ by $\gamma_{ab}$  and  changing the volume element from $d^4x$ to
$d^4x \sqrt{-\gamma}$ {\it but without} changing ordinary derivatives into covariant derivatives. The action
 again acquires a kinematic dependence on $\gamma_{ab}$ which we shall explicitly exhibit by writing it as 
 \begin{equation}
 A(\phi_A,\partial\phi_A,\eta_{ab})\to
 A_\partial(\phi_A,\partial\phi_A,\gamma_{ab})=\int d^4x \sqrt{-\gamma}\, L_\partial(\phi_A,\partial\phi_A,\gamma_{ab})
 \end{equation}
(The subscript $\partial$ in $A_\partial$ is to remind ourselves that, in obtaining this action, ordinary derivatives 
are retained as they were.)
 We can again obtain a second rank symmetric tensor $\mathcal{S}_{ab}$ by taking the functional derivative of the action  with respect to $\gamma_{ab}$ and then setting $\gamma_{ab}=\eta_{ab}$:
\begin{equation}
\delta A_\partial =  \frac{1}{2} \int d^4 x\, \sqrt{-\gamma} \, \mathcal{S}^{ab} \delta \gamma_{ab} ; \quad
\mathcal{S}^{ab}(x)\equiv \left[\frac{2}{\sqrt{-\gamma}}
\frac{\delta A_\partial}{\delta \gamma_{ab} (x)}\right]_{\gamma=\eta}
\label{deftabours1}
\end{equation}
More explicitly
\begin{equation}
\mathcal{S}^{ab}(x)\equiv \left[
\frac{2}{\sqrt{-\gamma}} \frac{\partial L_\partial\sqrt{-\gamma}}{\partial \gamma_{ab} } \right]_{\gamma=\eta}
=2\left[
 \frac{\partial L\sqrt{-\gamma}}{\partial \gamma_{ab} } \right]_{\gamma=\eta}
\label{deftabours}
\end{equation}
This procedure  provides another possible prescription for obtaining $\mathcal{S}^{ab}$ 
in flat spacetime. Once again, the $\gamma_{ab}$ is purely a bookkeeping device and, in the end of the calculations, we shall set $\gamma_{ab}=\eta_{ab}$.

Both $\mathcal{S}^{ab}$ and $T^{ab}$ are Lorentz invariant tensors.
In general, there are two crucial differences between these two tensors
$T^{ab}$ and $\mathcal{S}^{ab}$: (i) The $T^{ab}$ is obtained from a generally covariant lagrangian and hence is generally covariant; the $\mathcal{S}^{ab}$ {\it need not} be generally covariant (ii) The $T^{ab}$ satisfies the identity $\nabla_a T^{ab}=0$ since it arises from an action which is a generally covariant scalar; the $\mathcal{S}^{ab}$  {\it need not} satisfy  this identity.  Having said these, one must note that these two tensors are identical whenever the action does not depend on the derivatives of the metric. For spin-0 field, spin-1 field and for a relativistic particle,  the generally covariant action 
in Eq.(\ref{gencovact}) is independent of the derivatives  $\partial_a \gamma_{bc}$ of the metric tensor. Hence, in all these three --- physically important cases ---
the two definitions lead to identical energy momentum tensors: $T^{ab}=\mathcal{S}^{ab}$. Even our apparently non-covariant definition will lead to a generally covariant energy momentum tensor which satisfies the condition  $\nabla_a \mathcal{S}^{ab}=0$. 

For the spin-2 field these two definitions differ.  To find $\mathcal{S}^{ab}$, we need to start with the action in the form
\begin{eqnarray}
A_\partial &=&\frac{1}{4} \int d^4x\, \sqrt{-\gamma}\,  \partial_a h_{bc}\partial_i h_{jk} \, M^{abcijk}(\gamma^{mn})\nonumber\\
&=& \frac{1}{4} \int d^4x\, \sqrt{-\gamma}\,  \partial_a h_{bc}\partial_i h_{jk} 
\left[ \gamma^{ai}\gamma^{bc}\gamma^{jk}
 -\gamma^{ai}\gamma^{bj}\gamma^{ck}
+2 \gamma^{ak}\gamma^{bj}\gamma^{ci}
 -2\gamma^{ak}\gamma^{bc}\gamma^{ij}\right]
\label{habincstour}
\end{eqnarray}
which differs from that in Eq.(\ref{habincst}) by the fact that we have {\it not} changed $\partial_a$'s to
$\nabla_a$'s. The tensor $\mathcal{S}^{ab}$ can now be calculated using Eq.(\ref{deftabours}):
\begin{equation}
\mathcal{S}^{pq}(x)\equiv\left[
\frac{2}{\sqrt{-\gamma}}  \frac{\partial L\sqrt{-\gamma}}{\partial \gamma_{ab} } \right]_{\gamma=\eta}
=\frac{1}{2}\left[ \frac
{\partial \sqrt{-\gamma}M^{abcijk}(\gamma^{mn})}{\partial \gamma_{pq} } \right]_{\gamma=\eta}\partial_a h_{bc}\partial_i h_{jk}
\label{taboursforh}
\end{equation}
Again it is possible to write down the explicit expression for this but, fortunately, we will not need it. But it is
obvious from this expression that this tensor is quadratic in $\partial_a h_{bc}$ and does \emph{not}
involve second derivatives of $h_{ab}$. We shall see in the next section that  it is this object which
governs, through a term $\mathcal{S}^{ab}h_{ab}$ in the Lagrangian, the coupling of gravity to itself at the lowest nontrivial order. 

\section {Sneak Preview: Reverse engineering of Einstein's theory}

We want to obtain Einstein's theory by starting from the action for a spin-2 field in flat spacetime and
coupling it to some kind of energy momentum tensor for $h_{ab}$ and iterating the process. Given the
ambiguities in the definition of the energy momentum tensor described in the last section, it makes sense to do it the other way round and  identify the correct form of the tensor to which $h_{ab}$ couples to
in Einstein's theory. This exercise is straight forward: (i) Start with an action functional
$A_g[g_{ab}]$ which leads to Einstein's field equations for the metric tensor $g_{ab}$. (ii) Define the spin-2 field 
through $g_{ab}=\eta_{ab}+\lambda h_{ab}$, where $h_{ab}$ has the dimension (length)$^{-1}$ and 
$\lambda$ has the dimensions of length. (iii) Substitute $g_{ab}=\eta_{ab}+\lambda h_{ab}$ in $A_g[g_{ab}]$
and expand in a Laurent-Taylor series in $\lambda$ [or, which is the same thing, do a functional Taylor
series in $h_{ab}$]. Now, if our ideas are correct, two things must happen: (a) the lowest order term should give the action functional for spin-2 field in flat spacetime with a suitable choice for $\lambda$; (b) the next order term will have a lagrangian
of the form $\lambda h_{ab} K^{ab}$ and we should be able to read off $K^{ab}$. 
We will carry out this exercise and then comment on various issues.

The conventional action principle for general relativity is  the Einstein-Hilbert action given 
  by (with $\lambda^2=4\pi G$)
   \begin{equation}
   A_{\rm EH} \equiv \frac{1}{4\lambda^2} \int R\sqrt{-g} d^4x 
    \equiv \frac{1}{4\lambda^2} \int d^4x  [\sqrt{-g} L_{\rm quad}  - \partial_j P^j]\equiv A_{quad}+A_{sur}
\label{aehbegin}
   \end{equation}  
   where
    \begin{equation}
    L_{\rm quad} = g^{ab} \left(\Gamma^i_{ja} \Gamma^j_{ib} -\Gamma^i_{ab} \Gamma^j_{ij}\right)
    \label{lquad}
    \end{equation}
      and 
    \begin{equation}
    P^c = \sqrt{-g} \left(g^{ck} \Gamma^m_{km}- g^{ik} \Gamma^c_{ik}\right) 
   = \sqrt{-g} \, ( g^{ac} g^{ji}- g^{ia} g^{cj}) \partial_i g_{ac}
    \label{defpcone}
     \end{equation}
     The quantity $L_{quad}$ is what is usually called the $\Gamma^2$ lagrangian and is quadratic in the first derivatives of the metric. The term $\partial_iP^i$ integrates to a surface term and is usually ignored while deriving the field equations by assuming that ``suitable'' boundary conditions can be imposed. A more formal
     route is to add a suitable boundary term to cancel this, thereby essentially reducing the action to one based on $L_{quad}$.
   
 Classically, there is no way of deciding whether $A_{EH}$ or $A_{quad}$ is the ``correct" action, since both 
 lead to the same field equations. Let us first consider   $A_{quad}$ and determine to what second rank tensor it self-couples at the lowest order:
      Since $\Gamma\simeq \partial g$, one can express $A_{quad}$ as a quadratic expression 
     in $\partial_a g_{bc}$. 
  Straightforward algebra gives this to be of the form:
 \begin{eqnarray}
A_{quad}
&=& \frac{1}{4\lambda^2} \int d^4x\, \sqrt{-g}\,  \partial_a g_{bc}\partial_i g_{jk} 
\left[ g^{ai}g^{bc}g^{jk}
 -g^{ai}g^{bj}g^{ck}
+2 g^{ak}g^{bj}g^{ci}
 -2g^{ak}g^{bc}g^{ij}\right]\nonumber\\
&=& \frac{1}{4\lambda^2} \int d^4x\, \sqrt{-g}\,  \partial_a g_{bc}\partial_i g_{jk} \, M^{abcijk}(g_{mn})
\label{final}
\end{eqnarray} 
with the same functional form as the $M^{abcijk}$ defined in Eq.~(\ref{spintwo}) ! {\it This  is a miracle} and all the
results of this paper are essentially an exploitation of this miracle.  This result means that, if, after obtaining the flat spacetime action for spin-2 in   Eq.(\ref{spintwo}), we have (i) ``just replaced"  $\eta_{ab}$ by
$g_{ab}$ in $M^{abcijk}$ and $d^4x$ by $d^4x\sqrt{-g}$ and (ii) used $g_{ab}=\eta_{ab}+\lambda h_{ab}$
in the derivatives $\partial_a h_{bc}$, we would have got the correct Einstein's theory. All the (infinite) iterations are required only to understand why this is legal.

Let us now proceed with the original programme of reverse engineering this action to find out what it couples to at the lowest order.  This is quite straightforward.  Substituting $g_{ab}=\eta_{ab}+\lambda h_{ab}$ in Eq. (\ref{final}),
we get  $\partial_a g_{bc}\partial_i g_{jk}=\lambda^2  \partial_a h_{bc}\partial_i h_{jk}$ exactly. The Taylor
series expansion of $\sqrt{-g}M^{abcijk}$ gives:
\begin{equation}
\sqrt{-g}M^{abcijk}=M^{abcijk}(\eta^{mn})+\lambda\left[ \frac
{\partial \sqrt{-g}M^{abcijk}(g^{mn})}{\partial g_{pq} } \right]_{g=\eta}h_{pq}+{\cal O}(\lambda^2)
\end{equation}
Putting them together, we get the expansion:
 \begin{equation}
A_{quad}=\frac{1}{4} \int d^4x\,  \partial_a h_{bc}\partial_i h_{jk} \, M^{abcijk}(\eta^{mn})
+\frac{\lambda}{4} \int d^4x\,  \partial_a h_{bc}\partial_i h_{jk} \,
\left[ \frac
{\partial \sqrt{-g}M^{abcijk}(g^{mn})}{\partial g_{pq} } \right]_{g=\eta}h_{pq}+{\cal O}(\lambda^2)
\end{equation}
But the integrand of the second term contains precisely the quantity we defined in Eq. (\ref{taboursforh}). Using it, we get the final answer:
 \begin{equation}
A_{quad}=\frac{1}{4} \int d^4x\,  \partial_a h_{bc}\partial_i h_{jk} \, M^{abcijk}(\eta^{mn})
+\frac{\lambda}{2} \int d^4x\, \mathcal{S}^{pq}h_{pq}+{\cal O}(\lambda^2)
\label{proofone}
\end{equation}
with $\mathcal{S}^{pq}$ given by:
\begin{equation}
\mathcal{S}^{pq}
=\frac{1}{2}\left[ \frac
{\partial \sqrt{-\gamma}M^{abcijk}(\gamma^{mn})}{\partial \gamma_{pq} } \right]_{\gamma=\eta}\partial_a h_{bc}\partial_i h_{jk}
\label{needed}
\end{equation}
We, therefore, have proved that the coupling of gravity to itself, at least to the lowest order, is to a strange beast, defined in a noncovariant way. 
To $\mathcal{O}(\lambda)$,  the field $h_{ab}$
couples to a quantity which is quadratic in the first derivatives, $\partial_a h_{bc}$, of the field and does
not couple to an object which has second derivatives of the field. 
The standard  energy momentum tensor  $T^{pq}$ defined in Eq.~(\ref{deftab})  involves second derivatives
of the field and hence a naive coupling of the form $T^{pq} h_{pq}$ will not match with what we have
found by explicit computation above. 
(The  proof in Eq.~(\ref{proofone}), of course, gives the result only to the lowest order and one needs to know
whether gravity consistently couples to $\mathcal{S}^{ab}$ [as defined by Eq.~(\ref{taboursforh})] to all orders in the coupling
constant. It should be intuitively obvious that it will,  but we shall provide an explicit proof in Section IV A.)

It is, anyway, easy to understand that the self-coupling term in gravitational lagrangian {\it cannot} be of the form  $\lambda h_{ab}T^{ab}_G$  where $T^{ab}_G$ is the energy momentum tensor of the graviton. The reason is the following:
A term in the lagrangian proportional to $\lambda h_{ab}T^{ab}$ where $T^{ab}$ is due to {\it external} matter fields (assumed to be independent of $h_{ab}$ to this order), will lead to the equation of motion of the type $\partial^2 h=\lambda T$. Hence a coupling of the type
 $\lambda h_{ab}T^{ab}$ is equivalent to requiring the source being $T_{ab}$. Consider now the coupling of gravity to itself through a term of the type $\lambda h_{ab}t^{ab}(h)$ where $t_{ab}$ is some tensor which explicitly depends on the graviton field $h_{ab}$. When this term is varied with respect to $h_{ab}$ to get the equations of motion we will obtain {\it two} terms:
$t^{ab}+(\partial t_{ij}/\partial h_{ab})h^{ij}$ both of which will act as a source to gravity at next order. If we want the source to be the energy momentum tensor of graviton field, $T_{ab}^G$, then the coupling {\it cannot} be of the form
$h_{ab}T^{ab}(h)_G$ since this will lead to the wrong source. What we find is that the coupling in the lagrangian should be to ${\cal S}^{ab}$ if the source of the gravity is to be the energy momentum tensor to the lowest order. (In this paper, we shall use the terminology, ``$A$ is couples to be $B$", if the lagrangian at the relevant context has the term $AB$.
This does not necessarily mean that $B$ acts as a source term in Euler Lagrange equations when $A$ is varied, since --- in general --- $B$ could be a functional of $A$.)

So far we have obtained the lowest order self-coupling from the $\Gamma^2$ action and we will now turn our attention to the surface term.  We will  now prove a  strong result: {\it It is impossible to obtain the Einstein-Hilbert action, especially the  $A_{sur}$ term, by starting from an action for spin-2 field which is quadratic
in the first derivatives and iterating in powers of $\lambda$}. To see this qualitatively, note the structure of the Taylor series expansion for $A_{sur}$; symbolically:
\begin{equation}
A_{sur}\sim\frac{1}{\lambda^2}\int d^4x\, \partial [\lambda\partial h (\eta +\lambda h +\lambda^2 h^2 
+\mathcal{O}(\lambda^3))]
\sim\frac{1}{\lambda}\int d^4x\,  \partial [\eta\, \partial h] + \int d^4x\,  \partial [h\, \partial h ] +\lambda \int d^4 x \, \partial [h^2  \partial h]
+\mathcal{O}(\lambda^2)
\label{expa}
\end{equation}
 (The Lagrangian in $A_{\rm sur}$
 is linear in first derivative of $g_{ab}$. The substitution $g_{ab} = \eta_{ab} +\lambda h_{ab}$
 will lead to a $\lambda \partial h$ type term which  --- on multiplication by the pre-factor
 $\lambda^{-2}$ --- will give rise to the lowest order term which scales as $\lambda^{-1}$).
The  third term of the above expansion, which is $\mathcal{O} (\lambda)$
 and  has the structure $\partial(h^2\partial h)$, can be combined with the coupling term $(\lambda/2) \mathcal{S}^{pq} h_{pq}$ in Eq.~(\ref{proofone}) which is also $\mathcal{O}(\lambda)$. 
 If we  write:
 \begin{equation}
 \partial(h^2\partial h)\sim h[h\partial^2 h +(\partial h)^2]
 \label{futile}
 \end{equation}
 it might seem that the 
 $\mathcal{S}^{pq}$ in the coupling term $(\lambda/2) \mathcal{S}^{pq} h_{pq}$ in Eq.~(\ref{proofone}) changes with a contribution from the second derivatives of the field. One might wonder whether this will 
 help us to get $A_{sur}$ by using the coupling $(\lambda/2) T^{pq} h_{pq}$ and exploiting the second derivatives $\partial^2h$ in $T^{pq}$. {\it This idea, however, will not work}. 
 The term we need [$\mathcal{O}(\lambda)$ term] is the {\it third} term in the Taylor series expansion
 in Eq.(\ref{expa}) and to get $A_{EH}$ as the final answer, we {\it must} obtain first two terms as well.
 The two
 {\it leading terms} in
 Eq.~(\ref{expa}) are of $\mathcal{O} (1/\lambda)$ and $\mathcal{O} (1)$ and hence  cannot be obtained by
 the iterative process by coupling in ascending powers of $\lambda$.
 
 More explicitly,  the two leading   terms in $A_{sur}$ in the Einstein-Hilbert Lagrangian are:
 \begin{eqnarray}
A_{\rm sur}&=&
-\frac{1}{4\lambda}\int d^4x\,  \partial_a \partial_b[h^{ab}-\eta^{ab}h^i_i] \nonumber\\
&+&\frac{1}{4}\int d^4x\,  \partial_c \left[ \frac{1}{2} h \partial^c h- \frac{1}{2} h \partial_k h^{kc}
  -h^{kc} \partial_k h - h^{ab} \partial^c h_{ab} + h^{bc} \partial_k h_{b}^k + h^{ak}\partial_kh_{a}^c\right]
+\mathcal{O}(\lambda)
\label{ehtaylor}
  \end{eqnarray}
 Thus the leading term in Einstein-Hilbert Lagrangian (the term in the first line of the above equation) 
\begin{equation}
A_{sur}\approx
-\frac{1}{4\lambda}\int d^4x\,  \partial_a \partial_b[h^{ab}-\eta^{ab}h^i_i]
\label{leading}
\end{equation}
is non analytic in $\lambda$ when expanded in  
 $g_{ab} = \eta_{ab} +\lambda h_{ab}$. Note that one cannot get out of this non analyticity
 by cheap tricks  like rescaling $h_{ab}$ to $\lambda h_{ab}$. The dimension of the genuine
 graviton field and the form of the  zeroth order lagrangian uniquely fixes the scaling of $h_{ab}$ and there is no freedom
 for dimensionful scaling left.  
(In fact, it turns out that this non-analyticity is vital for the interpretation of the
 surface term as horizon entropy in semiclassical gravity; so it is not a trivial issue. This is discussed in detail in a separate publication \cite{nonanalytic} ) For us, the importance of the non-analyticity lies in the following fact:
 If one starts with the quadratic  spin-2 graviton action and iterate with self coupling,
 it is impossible to obtain $R\sqrt{-g}$, since it requires obtaining a piece non-analytic in $\lambda$.
One may wonder how previous ``derivations" obtained this piece. This was added by hand; since all the terms in
$A_{sur}$ are four divergences, any part of it can be added by hand without affecting the equations of motion
and this is what was done. (We will discuss this
in detail, in the context of the derivation by Deser \cite{deser} in Section IV C.)

The  Lagrangian in the leading order surface term in Eq. (\ref{leading}), viz.,  $L_{sur}\equiv \partial_a \partial_b[h^{ab}-\eta^{ab}h^i_i]$, which has 
 some interesting properties. First, it {\it is} gauge invariant under the transformation Eq.(\ref{gt}),
 as can be easily checked. (This, of course, means that one cannot set this term zero by a gauge choice).
 Second, it provides a simple counter example to the belief that if a functional is invariant under infinitesimal
 gauge transformations [that is, under Eq.(\ref{gt}), with $\xi$ treated as first order infinitesimal], then
 the expression will be generally covariant. This belief originates from the fact metric tensor transforms like
 in Eq.(\ref{gt}) under infinitesimal coordinate transformations 
 $(x^i \to x^i + \xi^i(x))$
 and one thinks of [erroneously] the finite
 coordinate transformations as arising from ``exponentiating" the infinitesimal ones.  
 Explicitly, the functional
 \begin{equation}
  F = \partial_i \left[ \sqrt{-g} \, \partial_l g_{jk} \left( g^{jk} g^{li} - g^{lk} g^{ji}\right) \right]
  \label{fourdiv}
  \end{equation}
  is clearly not invariant under arbitrary coordinate transformations. But if we take $g_{ab} = \eta_{ab}
  + \lambda h_{ab}$, then the expression for $F$ becomes, to linear order in $h$,
    \begin{equation}
  F_{\rm lin} \approx  \lambda \partial_i \left[ \partial_l h_{jk} \left( \eta^{jk} \eta^{li} - \eta^{lk} \eta^{ji}\right) \right]
  =-\lambda \partial_i  \partial_l\left[  h^{il} -\eta^{il} h^j_j \right]
  \end{equation}
  which is the same as the integrand in Eq.~(\ref{leading}).
  This expression, however, \emph{is} gauge invariant under the transformations in Eq.~(\ref{gt}). This shows that
  it is possible to have scalars which are invariant under infinitesimal gauge transformations but
  are not generally covariant.

\section{ From gravitons to gravity: general procedure}

After the sneak preview of the results to come, we shall now return to the original task of coupling the
spin-2 field to matter, as well as to itself, self consistently to all orders. We shall start with the issue of coupling
the spin-2 field to \emph{other} matter fields self-consistently to all orders to see   how an externally specified, $h_{ab}(x)$ affects the dynamics of $\phi_A(x)$.

Consider a field $\phi_A(x^a)$ described 
by a Lagrangian density $L(\phi_A,\partial\phi_A,\eta_{ab})$
in {\it flat} spacetime, in the Cartesian coordinates in which the metric is $\eta_{ab} = $ dia $(-1, 1,1,1)$. The index $A$ formally denotes all the indices the field carries depending on its spin; we will assume the field is bosonic for simplicity.  To couple $h_{ab}$ to this matter field, we need to first find a suitable, second rank tensor field $K^{ab}$, defined in terms of the matter variables. In Section II,  we introduced two such tensors (among infinite number of possibilities) in Eq.~(\ref{deftab}) and Eq.~(\ref{deftabours}). Both can be generically expressed as functional derivatives in the form:
\begin{equation}
\delta A_0 =  \frac{1}{2} \int d^4 x\, \sqrt{-\gamma} \, K^{ab} \delta \gamma_{ab} ; \quad
K^{ab}(x)\equiv \left[\frac{2}{\sqrt{-\gamma}}
\frac{\delta A_0}{\delta \gamma_{ab} (x)}\right]_{\gamma=\eta}
\label{deftabalt}
\end{equation}
If we use $A_0=A_\nabla(\phi_A,\nabla\phi_A,\gamma_{ab})$ (with covariant derivatives), we get the conventional
energy momentum tensor $K^{ab}=T^{ab}$. If, instead, we use $A_0=A_\partial(\phi_A,\partial\phi_A,\gamma_{ab})$
we get $K^{ab}=\mathcal{S}^{ab}$ which is a hybrid object,  except for the relativistic particle, spin-0 field or spin-1 field
for which the definitions coincide and $T^{ab}=\mathcal{S}^{ab}$. For most part of our algebra below, we need not specify whether we are using $A_\nabla$ or $A_\partial$ and we will use the generic symbols $A$ and $K^{ab}$
to stand for either definition.

We now want to couple the second rank symmetric tensor field $h_{ab}$   to $K^{ab}$. To the lowest order, this is done by changing the action from $A_0$ to $A_{\leq 1}\equiv A_0+A_1$ where
$A_1$ is chosen such that:
\begin{equation}
\delta A_1=\frac{\lambda}{2} \int d^4x_1 \sqrt{-\gamma}\, K^{ab}(x_1)\delta h_{ab}(x_1)
=\lambda\int d^4x_1\left[\frac{\delta A_0}{\delta \gamma_{ab} (x_1)}\right]_{\gamma=\eta}  \delta h_{ab}(x_1)
\label{start}
\end{equation}
where $\lambda$ is a coupling constant.
To the lowest order, $K^{ab}$ is independent of $h_{ab}$ and we can integrate this  to obtain the action:
\begin{equation}
A_{\leq 1}= A_0+A_1=
A_0 +\lambda\int d^4x_1\left[\frac{\delta A_0}{\delta \gamma_{ab} (x_1)}\right]_{\gamma=\eta} h_{ab}(x_1)
=A_0 +\frac{\lambda}{2} \int d^4x_1 \sqrt{-\gamma}\,  K^{ab}(x_1)h_{ab}(x_1)
\label{aone}
\end{equation}
The addition of this coupling will, however, change the definition of $K^{ab}$, since the second term
$A_1$  contributes to $K^{ab}$ via
Eq.({\ref{deftabalt}). To take this into account, we need to add a term $A_2$ in a manner similar to what we did in Eq.(\ref{start}); that is, we need to impose:
\begin{equation}
\delta A_2
=\lambda\int d^4x_2\left[\frac{\delta A_1}{\delta \gamma_{cd} (x_2)}\right]_{\gamma=\eta}\delta h_{cd}(x_2)
\end{equation}
Using the form for $A_1$ in Eq.(\ref{aone}) we get
\begin{equation}
\delta A_2=
\lambda^2 \int d^4x_1 d^4x_2 \left[\frac{\delta^2 A_0}{\delta\gamma_{cd}(x_2)
\delta\gamma_{ab}(x_1)}\right]_{\gamma=\eta}h^{ab}(x_1) \delta h^{cd}(x_2)
\label{startagain}
\end{equation}
Integrating, we get the second order correction to be
\begin{equation}
A_2=
\frac{\lambda^2}{2} \int d^4x_1 d^4x_2 \left[\frac{\delta^2 A_0}{\delta\gamma_{cd}(x_2)
\delta\gamma_{ab}(x_1)}\right]_{\gamma=\eta}h^{ab}(x_1) h^{cd}(x_2)
\end{equation}
 It is obvious that this term will bring about another correction etc. The sum of the infinite series of terms
 in the action  will be
\begin{equation}
A_\infty=\sum_{n=0}^\infty \frac{\lambda^n}{n!}\int d^4x_1 \ldots d^4x_n
\left[
\frac{\delta^n A_0}{\delta\gamma_{ab}(x_1)....\delta\gamma_{ij}(x_n)}\right]_{\gamma=\eta} 
h_{ab}(x_1)......h_{ij}(x_n)
\end{equation}
which is just a functional Taylor series expansion leading to:
\begin{equation}
A_\infty=A_0(\gamma_{ab}+ \lambda h_{ab})\,\big|_{\gamma=\eta}
\label{cstone}
\end{equation}
An alternative way of obtaining the same result is note that, at every order, we have the recurrence
relation, similar to Eq.~(\ref{start}) and Eq.~(\ref{startagain}):
\begin{equation}
\delta A_{n+1}
=\lambda\int d^4x_1\left[\frac{\delta A_n}{\delta \gamma_{ab} (x_1)}\right]_{\gamma=\eta}\delta h_{ab}(x_1)
\end{equation}
which is the same as:
\begin{equation}
\frac{\delta A_{n+1}}{\lambda\delta h_{ab} (x)}=\left[\frac{\delta A_n}{\delta \gamma_{ab} (x)}\right]_{\gamma=\eta}
\end{equation}
Summing both sides to to infinite orders, we find that $A_\infty$ satisfies the relation
\begin{equation}
\frac{\delta A_\infty}{\lambda\delta h_{ab} (x)}=\left[\frac{\delta A_\infty}{\delta \gamma_{ab} (x)}\right]_{\gamma=\eta}
\label{cons}
\end{equation}
which has the general solution given by Eq.(\ref{cstone}). This is our key result.

Thus we can consistently couple a field $h_{ab}$ to $K^{ab}$ by the rule given in Eq.~(\ref{cstone}):
This allows us to compute the effect of an external $h_{ab}$ on the system if we insist that 
the external field consistently couples 
to a tensor $K^{ab}$ which can be expressed as in Eq.~(\ref{deftabalt}).
Since the curvilinear metric $\gamma_{ab}$
was introduced only as bookkeeping device to allow for variation of the action,  the final
action is given by replacing $\gamma_{ab}$ by $\eta_{ab}$ at the end of the calculation.
Note that there is  subtle difference between Eq.(\ref{cstone}) and the expression obtained
by replacing $\eta_{ab}$ by $(\eta_{ab}+ \lambda h_{ab})$ in the original action.
The latter one will miss, for example,  the $\sqrt{{\rm det}|\gamma|} \to \sqrt{{\rm det}|\eta +h|}$ kind of factors.
 We need to introduce $\gamma_{ab}$
in order to provide a placeholder in the final expression. 

Here comes the parting of ways. If we had chosen $A=A_\nabla(\phi_A,\nabla\phi_A,\gamma_{ab})$ (with covariant derivatives), then $K^{ab}=T^{ab}$ is the standard energy momentum tensor and our result gives the final matter action to be:
\begin{equation}
A_\infty=A_0(\phi_A,\nabla_{(g)}\phi_A, g_{ab})
\end{equation}
where we have used the abbreviation $g_{ab}\equiv \eta_{ab}+ \lambda h_{ab}$ and $\nabla$ is defined with respect to this metric. This is a generally covariant
matter action in the spacetime with metric $g_{ab}$ and agrees with all the text book results. It should, however, be stressed that this cannot be considered a {\it derivation} of general covariance of matter action when self-consistently coupled to spin-2 field. This is because we made a rule for finding $T^{ab}$ which has general covariance with
respect to curved spacetime built in as an {\it assumption}. All that has been shown is that
this extends to an interpretation of $g_{ab}$ as a metric tensor and only the combination $g_{ab}\equiv \eta_{ab}+ \lambda h_{ab}$ is relevant for matter sector. 

Since the final matter Lagrangian we obtained \emph{is}
generally covariant, its energy momentum tensor has zero covariant divergence, leading to:
\begin{equation}
\nabla_a\left[\frac{\delta A_\infty}{\delta g_{ab}}\right]=0
\label{coderoftab}
\end{equation}
where $\nabla$ is the covariant derivative operator corresponding to $g_{ab}=\eta_{ab}+ \lambda h_{ab}$. 

On the other hand, if we had chosen $A=A_\partial(\phi_A,\partial\phi_A,\gamma_{ab})$ (without covariant derivatives), then $K^{ab}=\mathcal{S}^{ab}$ is the beast we have introduced in Section II D
and our result gives the final matter action to be:
\begin{equation}
A_\infty=A_0(\phi_A,\partial\phi_A, g_{ab})
\end{equation}
In general, this is {\it not} a 
 a generally covariant
 matter Lagrangian since we have not replaced partial derivatives by covariant derivatives. The metric appears only
through $\sqrt{-g}$ factor and by the replacement of $\eta_{ab}$ by $g_{ab}$. This, of course,
does not matter for the lagrangians  of relativistic particle, spin-0 field or spin-1  field.  In all these cases this
 prescription \emph{does} lead to a generally covariant lagrangian though this is {\it not } by design.
 The Eq.(\ref{coderoftab}) also holds for the 
spin-0 or spin-1 field or for the  relativistic particles {\it but not in general}. (We shall comment on this 
 in Section V.)

\subsection{How to obtain Einstein gravity from the spin-2 field}

To see where all this is leading to, let us consider next the real issue: that of coupling the graviton field to itself.
 In our approach this is ridiculously simple. 
We merely use the fact that
the analysis leading to Eq.~(\ref{cstone}) 
was completely independent of the form of $A_0$ as well as the nature of the fields
$\phi_A$. Hence we can use the same prescription when $\phi_A$ is the second 
rank symmetric tensor field $h_{ab}$ itself. 

The key ambiguity, of course, is whether we want to use $A_\nabla$ and couple to $T^{ab}$ or use
$A_\partial$ and couple to $\mathcal{S}^{ab}$. Let us use the second procedure first: The $A_\partial$ for the
graviton field is obtained from the first line of  Eq.~(\ref{spintwoact})  by 
 replacing 
$\eta^{ab}$ by $\gamma^{ab}$ in 
$M^{abcijk}(\eta)$ and multiplying 
by $\sqrt{-\gamma}$. This is given by Eq.(\ref{habincstour}):
\begin{eqnarray}
A_\partial &=&\frac{1}{4} \int d^4x\, \sqrt{-\gamma}\,  \partial_a h_{bc}\partial_i h_{jk} \, M^{abcijk}(\gamma^{mn})\nonumber\\
&=& \frac{1}{4} \int d^4x\, \sqrt{-\gamma}\,  \partial_a h_{bc}\partial_i h_{jk} 
\left[\gamma^{ai} \gamma^{bc}\gamma^{jk}
 -\gamma^{ai}\gamma^{bj}\gamma^{ck}
+2\gamma^{ak} \gamma^{bj}\gamma^{ci}
 -2\gamma^{ak}\gamma^{bc}\gamma^{ij}\right]
\label{habincstour1}
\end{eqnarray}
Our prescription now requires that $A_\infty$ for the field $h_{ab} $ is obtained by replacing $\gamma_{ab}$
by $g_{ab} \equiv \eta_{ab} +\lambda h_{ab}$. This leads to the action 
\begin{eqnarray}
A_\infty &=&\frac{1}{4\lambda^2} \int d^4x\, \sqrt{-g}\,  \partial_a g_{bc}\partial_i g_{jk} \, M^{abcijk}(g^{mn})\nonumber\\
&=& \frac{1}{4\lambda^2} \int d^4x\, \sqrt{-g}\,  \partial_a g_{bc}\partial_i g_{jk} 
\left[ g^{ai}g^{bc}g^{jk}
 -g^{ai}g^{bj}g^{ck}
+2 g^{ak}g^{bj}g^{ci}
 -2g^{ak}g^{bc}g^{ij}\right]
\label{fiveeight}
\end{eqnarray}
which is precisely the $\Gamma^2$ action in general relativity.
The variation of this action will lead to Einstein's field equations {\it in vacuum} and we have achieved our goal.  

As an aside, we note that the above approach will work with any generic action functional
for spin-2 field. A different choice of the spin-2 action functional will lead to a different final theory. This aspect is briefly discussed in Appendix B since it is irrelevant to the main issues of this paper.

\subsection{The failure of the conventional procedure to lead to Einstein's theory}

Let us see what happens if we follow the conventional procedure. For this, we need to start with the 
$A_\nabla$ of graviton action and couple to the standard $T^{ab}$. This action is given by Eq. (\ref{habincst}):
\begin{eqnarray}
A_\nabla &=&\frac{1}{4} \int d^4x\, \sqrt{-\gamma}\,  \nabla_a h_{bc}\nabla_i h_{jk} \, M^{abcijk}(\gamma^{mn})\nonumber\\
&=& \frac{1}{4} \int d^4x\, \sqrt{-\gamma}\,  \nabla_a h_{bc}\nabla_i h_{jk} 
\left[ \gamma^{ai}\gamma^{bc}\gamma^{jk}
 -\gamma^{ai}\gamma^{bj}\gamma^{ck}
+2 \gamma^{ak}\gamma^{bj}\gamma^{ci}
 -2\gamma^{ak}\gamma^{bc}\gamma^{ij}\right]
\label{habincstalt}
\end{eqnarray}
Our prescription now requires that $A_\infty$  is obtained by replacing $\gamma_{ab}$
by $g_{ab} \equiv \eta_{ab} +\lambda h_{ab}$. This leads to the action 
\begin{eqnarray}
A_\infty &=&\frac{1}{4} \int d^4x\, \sqrt{-g}\,  \nabla_a h_{bc}\nabla_i h_{jk} \, M^{abcijk}(g^{mn})\nonumber\\
&=& \frac{1}{4} \int d^4x\, \sqrt{-g}\,  \nabla_a h_{bc}\nabla_i h_{jk} 
\left[ g^{ai}g^{bc}g^{jk}
 -g^{ai}g^{bj}g^{ck}
+2 g^{ak}g^{bj}g^{ci}
 -2g^{ak}g^{bc}g^{ij}\right]
\label{finalagain}
\end{eqnarray}
where the $\nabla$ operator is with respect to the metric $g_{ab}$. Now, since $\nabla_i g_{jk}=0$, we get:
\begin{eqnarray}
\nabla_a h_{bc}&=&\frac{1}{\lambda}\nabla_a[g_{bc}-\eta_{bc}]=-\frac{1}{\lambda}\nabla_a\eta_{bc}
=-\frac{1}{\lambda}[\Gamma_{ba}^i\eta_{ic}+\Gamma_{ca}^i\eta_{ib}]\nonumber\\
&=&-\frac{1}{\lambda}[g^{pi}\Gamma_{pba}(g_{ic}-\lambda h_{ic})+(b \leftrightarrow c)]
=-\frac{1}{\lambda}[\Gamma_{pba}(\delta^p_c-\lambda g^{pi} h_{ic})+(b \leftrightarrow c)]
\end{eqnarray}
and a similar expression for $\nabla_i h_{jk}$. Since these are multiplied by  $M^{abcijk}(g^{mn})$
which is  symmetric in $(b,c),(i,j)$ we can 
ignore the $(b\leftrightarrow c)$ term etc.
 We thus obtain
\begin{equation}
  \nabla_a h_{bc}\nabla_i h_{jk} \, M^{abcijk}(g^{mn})=\frac{4}{\lambda^2}[
\Gamma_{pba}\Gamma_{qki}(\delta^p_c-\lambda g^{pi} h_{ic})(\delta^q_j-\lambda g^{lq} h_{lj})]M^{abcijk}(g^{mn})
\label{sixtwo}
\end{equation}
Of the four terms which arise on expanding out the product, 
$(\delta_c^p - \lambda g^{pi} h_{ic})(\delta^q_j-\lambda g^{lq} h_{lj})$, 
the first term can be transformed, again using the symmetry of $M^{abcijk}(g^{mn})$ in $(b,c),(i,j)$ to give:
\begin{equation}
 \frac{4}{\lambda^2}\Gamma_{cba}\Gamma_{jki}M^{abcijk}(g^{mn})
= \frac{1}{\lambda^2}M^{abcijk}(g^{mn})[\Gamma_{cba}+\Gamma_{bca}][\Gamma_{jki}+\Gamma_{kji}]
= \frac{1}{\lambda^2}M^{abcijk}(g^{mn})\partial_a g_{bc}\,\partial_i g_{jk}
\end{equation}
This is precisely the lagrangian term in Einstein's theory, in the form of $\Gamma^2$ action 
[compare with Eq.~(\ref{fiveeight})]. Unfortunately, there are three more terms in Eq.~(\ref{sixtwo})
of the  $\Gamma\Gamma hh$ and
$\Gamma\Gamma h$ form. They do not vanish, they are not total divergences and they depend explicitly on
$h_{ab}$. Thus the action functional obtained by coupling the spin-2 field to the standard energy momentum tensor
is \emph{not} that of Einstein's theory.

The manner in which we led the reader to the result, it should not come as a surprise. We have, in fact, shown
in Section III explicitly that the standard action for Einstein's theory does  couple to 
$\mathcal{S}^{ab}$ in the lowest order. So clearly, we will not get Einstein's theory if we force feed the conventional energy momentum tensor $T^{ab}$ on to the spin-2 field.

\subsection{Comments on the previous work}

Finally, we shall discuss how the previous ``derivations'' escaped this problem.  As described in section I,
none of the previous derivations other than that of Deser \cite{deser} actually performs any iteration.
All the rest of them tacitly or explicitly brings in general covariance for the gravity sector, after which it is trivial
to obtain $R\sqrt{-g}$ as the Lagrangian; hence, we need not discuss them any further. Deser does perform the iteration using the first order form of 
gravity and using $\sqrt{-g}\, g^{ab} $ as the chosen variable. 
Let us first summarize this approach briefly in a slightly different language to bring the essential ingredients in to focus. 

We first note that the standard
Einstein-Hilbert action can be expressed in the first order Palatini form using the variables $\Gamma^a_{bc}$
and $f^{ab}\equiv \sqrt{-g}\, g^{ab}$ as:
\begin{equation}
A_{\rm EH}  = \frac{1}{4\lambda^2 } \int d^4x f^{ab} R_{ab} (\Gamma); 
\qquad f^{ab}\equiv \sqrt{-g}\, g^{ab}
\end{equation}
Varying $f^{ab}$ and $\Gamma^a_{bc}$ independently in this action will lead to standard Einstein's equations. If we substitute $f^{ab} = \eta^{ab} + \lambda q^{ab}$ into this action (without any approximations), then the lagrangian becomes:
\begin{equation} 
L_{\rm EH}  = \frac{1}{4 \lambda^2 }  [ (\eta^{ab} + \lambda q^{ab}) 
(R_{ab}^L (\Gamma)+  R_{ab}^Q(\Gamma))]
\label{deser1}
\end{equation}
where $R_{ab}^L (\Gamma)$ and $R_{ab}^Q (\Gamma)$ are the linear and quadratic parts of the 
Ricci tensor
\begin{equation}
R_{ab}^L (\Gamma) = \partial_c \Gamma^c_{ab} - \partial_b \Gamma_a; \quad R_{ab}^Q (\Gamma)=
\Gamma_c\Gamma^c_{ab} - \Gamma^c_{ad}\Gamma^d_{bc}; \quad \Gamma_c \equiv \Gamma^i_{ci}
\end{equation}
Expanding out the product in Eq.(\ref{deser1}), we get four terms which we will group as:
\begin{equation}
L_{EH}=L_0+L_1+L_2
=\frac{1}{4 \lambda^2 } \eta^{ab}R_{ab}^L (\Gamma) 
+\frac{1}{4 \lambda^2 } [\eta^{ab} R_{ab}^Q(\Gamma)+\lambda q^{ab}R_{ab}^L (\Gamma)]
+\frac{1}{4 \lambda^2 }\lambda q^{ab}R_{ab}^Q(\Gamma)
\label{deser2}
\end{equation}
We now notice something remarkable. 
\begin{itemize}
\item
 The first term $\eta^{ab}R^L_{ab} = \partial_i (\Gamma^i - \Gamma^{ij}_{\phantom{ij}j})$ is a total divergence. Let us assume we are allowed to drop this term. 
\item
 The second term has
 the piece
$[\eta^{ab} R_{ab}^Q(\Gamma)+\lambda q^{ab}R_{ab}^L (\Gamma)]$
which is essentially equivalent to the  zeroth order action for spin-2  graviton (in the second order formalism) {\it plus} a  \emph{very specific} four divergence term. [This will be apparent if we write
the term $ q R^L \sim q \partial \Gamma \sim \partial (q \Gamma) - \Gamma \partial q$; we find that it has a quadratic term plus \emph{a very specific} total divergence term $\partial (q \Gamma)$]. Let us assume we are allowed to start with this \emph{very specific} term as the lowest order graviton lagrangian.
\item
Granted these two wishes, we see that one can obtain from this $L_1$ term, the tensor $t_{ab}=(\delta L_1/\delta \eta^{ab})=(1/4 \lambda^2)
R_{ab}^Q(\Gamma)$ , purely formally. If we think of this as the energy momentum tensor for the graviton, then the next order
coupling should be $\lambda q^{ab}t_{ab}=(1/4 \lambda^2)[\lambda q^{ab}R_{ab}^Q(\Gamma)]$ which is precisely
the last piece $L_2$ in the Einstein-Hilbert lagrangian ! What is more, this last term is independent of $\eta^{ab}$
so $(\delta L_2/\delta \eta^{ab})=0$ and {\it no further iterations are required}. [Since we are working with
$\sqrt{-g}g^{ab}$ rather than $g^{ab}$, the variation actually gives $T_{ab}-(1/2)g_{ab}T$ rather than the energy momentum tensor itself; but this is irrelevant to our discussion.]
\end{itemize}
When one attempts to do this properly, one faces an important issue: As we have said before, one cannot really 
define things like $(\delta L/\delta \eta^{ab})$ where $\eta^{ab}$ is the Minkowski metric. So we first need to write
$L_1$ in curved spacetime with a metric  $\gamma^{ab}$ and compute variations with respect to this metric.
What do we do to the $\partial_a$ in the definition of $R_{ab}^L (\Gamma) = \partial_c \Gamma^c_{ab} - \partial_b \Gamma_a ?$ Let us suppose, we change them to $\nabla_a$ for the metric $\gamma^{ab}$.  Then the  $t_{ab}$
computed from the functional derivative will pick up additional terms. 
The action $L_1$ in curved spacetime is
\begin{equation} 
L_1 (\gamma)= \frac{1}{4 \lambda^2 }  [\sqrt{-\gamma}\gamma^{ab} R_{ab}^Q(\Gamma)+\lambda q^{ab}R_{ab}^L (\Gamma)];\qquad \bar\gamma^{ab}\equiv \sqrt{-\gamma}\gamma^{ab}
\end{equation}
where the $R_{ab}$ is now evaluated with partial derivatives replaced by covariant derivatives
with respect to $\gamma_{ab}$ etc.  The variation of this Lagrangian with respect to $\bar\gamma^{ab}$ gives
 (see eq. (8) of \cite{deser})
\begin{equation}
 t_{ab}  = \frac{1}{4\lambda^2} \left[ R_{ab}^Q (\Gamma) +\lambda
\sigma_{ab}\right]
\end{equation}
where the second term $\sigma_{ab}$ arises from the variation of
the $\partial_a\gamma_{bc}$ terms, because we have changed $\partial_a$ to $\nabla_a$. Its explicit form
\begin{equation}
\sigma_{ab} = \partial^c \left[ - \eta_{ab} \left(q_i^j \Gamma^i_{cj} - \frac{1}{2} q \Gamma_c\right)
- 2 q_c^i \Gamma_{(ab)i} - 2 q_{(a}^i \Gamma_{cib)} + 2 q_{(a}^i \Gamma_{b)ci} 
- q_{ab} \Gamma_c + 2 q_{c(a} \Gamma_{b)} \right]
\end{equation}
is irrelevant to us. We are now in trouble since the next order coupling $t_{ab}q^{ab}$ will
now have an unwanted term $q^{ab}\sigma_{ab}$,
in addition to the term we want (proportional to $q^{ab}R_{ab}^Q$).
Deser simply drops this term saying (see  his comment after eq. 9 of \cite{deser}):  {\it ``Note that we have not added the full $h^{\mu\nu}\tau_{\mu\nu}$, but rather used
the simple part of $\tau_{\mu\nu}$ only"} without any additional justification! Then, of course, one gets the $L_2$ as the
next term and iteration stops there.

It should now be obvious that Deser's derivation requires the following implicit assumptions: 
\begin{itemize}
\item
One should drop the $L_0$
term in Eq. (\ref{deser2}) unceremoniously, saying it is a total divergence [Deser does this in going from eq. (2) to eq. (4) in \cite{deser}]. It is precisely this term, which, in second order formalism, has
the non analytic behaviour $(1/\lambda)$ and is displayed as the first term in Eq. (\ref{ehtaylor}). Sure, it is a four divergence, but one can \emph{never} get it from graviton's
quadratic action and one needs to add and subtract this term, at will, to get Einstein-Hilbert action.

\item
One should start with $L_1$ which is not the graviton action that is quadratic in the derivatives of the field 
(that a  particle physicist would have written down from first principles) but
the one with a very specific total divergence added to it. This is precisely the $\mathcal{O}(1)$ in Eq. (\ref{ehtaylor}).
There is no way anyone could have guessed this specific total divergence term without knowing the 
final answer!

\item
One should drop the terms in $t_{ab}$ which arise from varying $\partial_a\gamma_{bc}$. But this is precisely the same as using our quantity $\mathcal{S}^{ab}$ ! Or, rather, not changing $\partial$ to $\nabla$ when one takes the
graviton action from flat spacetime to curved spacetime. So in real terms the two derivations match mathematically
and our conclusion stands: {\it Gravity self-couples to $\mathcal{S}^{ab}$; not to $T^{ab}$}. It just was not realized before.
\end{itemize}

If we are not attempting to {\it derive} Einstein's theory from the spin-2 theory but only want to {\it reinterpret} it in the 
field theoretical language, then one may be willing to live with the first two issues mentioned above. One can use the hindsight gained from general relativity and add and subtract four divergences at will to the action, to obtain the necessary final form. (We must then admit that the Venusian physicists whom Feynman keeps referring to in \cite{feynman} would never have got there.)
But the third issue is not a matter of opinion or point of view;  what quantity gravity couples to in becoming nonlinear
is a well defined mathematical question. Our analysis --- and proper interpretation of previous work --- gives an
unconventional answer.

\section{ Conclusions}

We have shown that it is not possible to obtain the Einstein-Hilbert {\it action} starting from the standard graviton action and iterating in the coupling constant. This is because of the existence of the total divergence term in the Einstein-Hilbert action which is non-analytic in the coupling constant, when expanded in terms of the graviton field. This result is crucial because a series of previous investigations \cite{tppapers}
have shown that the surface term is vital in the thermodynamics of horizons and in semiclassical gravity. In fact, I started this investigation to understand how the surface term --- and hence, possibly the entropy of horizons --- can be interpreted in terms of graviton field in a Minkowski background. The result shows that one simply cannot understand the surface term in a standard field theoretical language, using the graviton field. There is more to gravity than gravitons and this will be elaborated in a separate publication \cite{nonanalytic}.

In a strictly classical theory, what matters is the equation of motion and not the form of the action principle.
Hence, the fact that we can not get the surface term in Einstein-Hilbert action is not of concern if we are only interested in the Einstein's equations. Our analysis shows that it is indeed possible to
obtain the quadratic $\Gamma^2$ action (and thus the Einstein's equations) by starting from the the graviton action and iterating on the coupling constant. But to do this, we need to couple $h_{ab}$ to a second rank tensor $\mathcal{S}^{ab}$ which is different from from the standard energy-momentum tensor $T^{ab}_G$ of the graviton. Indeed, as we explained in Section III
(see the discussion after Eq.(\ref{needed})), if the {\it source} of gravity at each order of iteration has to be the energy-momentum tensor of the graviton evaluated at the previous order, then the {\it coupling} in the Lagrangian {\it cannot} be of the form $h_{ab}T^{ab}_G$  since the $h_{ab}$ dependence of the $T^{ab}_G$ will lead to an extra term on variation. A term in the lagrangian of the form $h_{ab}\mathcal{S}^{ab}$
does lead to the  energy-momentum tensor as the source of gravity. Identifying the nature of $\mathcal{S}^{ab}$
and bringing it into focus has been one of the results of this paper. 

If we were only interested in pure gravity, this would have been the whole story. But, in that case,
it is an unnecessary exercise. The {\it linear} spin-2 field, uncoupled to anything, is a perfectly consistent
theory and we need not try to couple it to itself. So the whole exercise has meaning only when we have both
matter and spin-2 field and we try to  couple them consistently. Then we need to assume that the spin-2 field couples to itself through $\mathcal{S}^{ab}$ while it couples to matter through $T^{ab}$. This assumption will lead consistently to Einstein's theory
and seems to be the most viable option, if we want to obtain standard gravity coupled to matter, starting from the graviton action. [Of course, in a world made of a spin-2 field coupled to matter made of {\it only}
relativistic particles, spin-0 fields and spin-1 fields, one can assume that all the coupling is through
$\mathcal{S}^{ab}$; this is because for matter made of these constituents, $\mathcal{S}^{ab}=T^{ab}$].

Two facts need to be borne in mind as regards this option. First, we do not know anything about the coupling of spin-2 field to itself except through standard gravity; and the analysis in Section III
shows that gravity does couple to itself through a term $h_{ab}\mathcal{S}^{ab}$. Second, there is no conflict with principle of equivalence even though the self-coupling term is $h_{ab}\mathcal{S}^{ab}$ while the
coupling to external source is through $h_{ab}T^{ab}$. What matters for principle of equivalence is the fact that
the source for gravity is always the energy-momentum tensor This is indeed assured in our approach and --- as have been stressed several times --- this requires a  self-coupling term of the form $h_{ab}\mathcal{S}^{ab}$ in the lagrangian.

\section*{Acknowledgements}
I thank S.Deser for his time and patience during {\it many} rounds of lively
email discussion, stretching over six months, regarding earlier versions of this draft which resulted
in considerable improvement as regards the contents and presentation. We
achieved fair amount of convergence but not in all issues.
 I thank Apoorva Patel. R. Nityananda and K.Subramanian for  
 detailed  comments on an earlier draft and D.Lynden-Bell, A.Lasenby, J.Bjorken for discussions on several
 related issues.

\section*{Appendix A: Physics of the spin-2 field - brief review}

In this Appendix we will briefly review the theory of the spin-2 field and collect together different results which are required later.  (This is done especially since I could not find a a convenient source for pedagogical 
discussion of spin-2 field; the results are somewhat scattered in the literature \cite{spintwo}). 

\subsection {Action functional for the spin-2 field}

The action for the non-interacting, massless, spin-2 field $h_{ab}$ is built out of scalars which are quadratic in
the derivatives $ \partial_a h_{bc}$. The most general expression will be the sum of different scalars obtained by
contracting  pairs of indices in $ \partial_a h_{bc}\partial_i h_{jk} $ in different manner. 
Since this product is symmetric in $(b,c)$ and $(j,k)$ and also under the interchange $(a,b,c) \to (i,j,k)$, 
it is easy to figure out  that, a priori,  seven different contractions are possible. For example, if $a$ is contracted with $i$, then there are two possibilities for contracting $b$ (with either $c$ or with $j$; contracting $b$ with $k$ is the same as
contracting $b$ with $j$).  These 
contractions will
lead to the terms 
$ c_1\partial_a h_{bc}\partial_i h_{jk}\eta^{ai}\eta^{bc}\eta^{jk}
+c_2 \partial_a h_{bc}\partial_i h_{jk}\eta^{ai}\eta^{bj}\eta^{ck} 
=c_1\partial_ah^b_b\partial^ah^j_j+
c_2\partial_ah_{bc}\partial^ah^{bc}
$ in the lagrangian with as yet undetermined constants $(c_1,c_2)$. 
For brevity, we will denote these two terms symbolically as $(ai,bc,jk),(ai,bj,ck)$.
Next, if  $a$ is contracted with $b$, there are again two inequivalent possibilities for contracting
 $c$ leading to $(ab,ci,jk),(ab,ck,ij)$. Finally if
 $a$ is contracted with $k$, there are three possible ways of contracting $b$ giving $(ak,bj,ci),(ak,bc,ij),(ak,bi,cj)$. 
 
Of these, the contraction $(ak,bc,ij)$ is the same as $(ab,ci,jk)$ since
$(ak,bc,ij)=(ic,jk,ab)$ under $(a,b,c) \leftrightarrow (i,j,k)$ and, of course, $(ic,jk,ab)=(ab,ci,jk)$. Similarly,
$(ak,bj,ci)=(ic,jb,ka) =(ib,jc,ka)$; the first equality comes from $(a,b,c) \leftrightarrow (i,j,k)$ symmetry while the 
second arises from $b \leftrightarrow c$ symmetry.
Since $(ib,jc,ka)=(ak,bi,cj)$ trivially, we need to retain only the first two out of the three possibilities in the last set. Thus 
dropping the two contractions $(ab,ci,jk)$ and $(ak,bi,cj)$ out of the 7 possibilities,
we are left with 5 different contractions: $(ai,bc,jk),(ai,bj,ck),(ab,ck,ij),(ak,bj,ci),(ak,bc,ij)$.  This will correspond to an
action for the spin-2 field of the form:
\begin{eqnarray}
A &=&\frac{1}{4} \int d^4x\,  \partial_a h_{bc}\partial_i h_{jk} 
\left[ c_1\eta^{ai}\eta^{bc}\eta^{jk}
 +c_2\eta^{ai}\eta^{bj}\eta^{ck}
 +c_3\eta^{ab}\eta^{ck}\eta^{ij}
+c_4\eta^{ak} \eta^{bj}\eta^{ci}
 +c_5\eta^{ak}\eta^{bc}\eta^{ij}\right]\nonumber\\
 &=& \frac{1}{4} \int d^4x\, \left[ 
c_1\partial_ah^b_b \partial^a h^j_j 
 +c_2 \partial_a h_{bc} \partial^a h^{bc}
+c_3 \partial_a h^{ab} \partial_i h^i_b
+c_4 \partial_a h_{bc} \partial^c h^{ba}
 +c_5\partial_a h^b_b \partial_i h^{ia}
 \right]
\label{spintwoact0}
\end{eqnarray}
Each term in the action in Eq.(\ref{spintwoact0}) is  of the kind
  $\partial_a h_{bc}\partial_i h_{jk} J^{abcijk}(\eta)$ where $J^{abcijk}$ is a cubic in $\eta^{lm}$ and hence
  is constant [i.e, all components are $0$ or $\pm 1$].  This allows one to ``swap" the derivatives $\partial_i$
  and $\partial_a$ by adding a total divergence, using the identity:
  \begin{equation}
  [\partial_a h_{bc}\partial_i h_{jk} -\partial_i h_{bc}\partial_a h_{jk}]J^{abcijk}
  =\partial_a [h_{bc}\partial_i h_{jk}(J^{abcijk}-J^{ibcajk})]
  \label{gen}
  \end{equation} 
Using this result, one can convert the $c_3$ term to the $c_4$ term and rewrite the action as
\begin{eqnarray}
A &=&\frac{1}{4} \int d^4x\,  \partial_a h_{bc}\partial_i h_{jk} 
\left[ c_1\eta^{ai}\eta^{bc}\eta^{jk}
 +c_2\eta^{ai}\eta^{bj}\eta^{ck}
 +(c_3+c_4)\eta^{ak} \eta^{bj}\eta^{ci}
 +c_5\eta^{ak}\eta^{bc}\eta^{ij}\right]+A_{\rm div}\nonumber\\
 &=& \frac{1}{4} \int d^4x\, \left[ 
c_1\partial_ah^b_b \partial^a h^j_j 
 +c_2 \partial_a h_{bc} \partial^a h^{bc}
+(c_3+c_4) \partial_a h_{bc} \partial^c h^{ba}
 +c_5\partial_a h^b_b \partial_i h^{ia}
 \right]+A_{\rm div} \equiv A_h + A_{\rm div}
\label{spintwoact1}
\end{eqnarray}
where
\begin{eqnarray}
A_{\rm  div}&=&
\frac{c_3}{4} \int d^4x\,  \partial_a h_{bc}\partial_i h_{jk} 
\left[ 
 \eta^{ab}\eta^{ck}\eta^{ij}
-\eta^{ak} \eta^{bj}\eta^{ci}
 \right]
=\frac{c_3}{4} \int d^4x\, 
\left[  \partial_a h^{ab} \partial_i h^i_b
-\partial_a h_{bc} \partial^c h^{ba}
\right]\nonumber\\
&=&\frac{c_3}{4} \int d^4x\,  
\partial_a[h^{ab}\partial_ih^i_b-h^{ib}\partial_ih^a_b]
\label{divamb}
\end{eqnarray}
which, being a total divergence, does not contribute to the equations of motion if suitable boundary conditions are imposed. 
Notice that there are {\it no} further ambiguities related to ``swapping" of derivatives
in the action in Eq.~(\ref{spintwoact0}); this is clearly not possible in 
the $c_1,c_2$ or $c_5$ terms, since the swapping leads to identical terms. 
Hence the only ambiguity is in the choice between
 the $c_3$ term for $c_4$ terms.

Interestingly enough, the
constants $c_1, c_2,  c_5, (c_3+c_4)$ in $A_h$ in  Eq.(\ref{spintwoact1}) can be determined except for an overall scaling by the requirement that the field equations should be  invariant under the gauge transformation:
\begin{equation}
 h_{ab}(x) \to h_{ab}(x) + \partial_a \xi_b (x) + \partial_b \xi_a (x).
 \label{gtapp}
 \end{equation}
This fixes the constants to be $c_1=-c_2=1; c_3+c_4=-c_5=2$ except for an overall scaling which 
is left as  (1/4) for future convenience. 
The resulting expression for the quadratic part of the action can be written in different forms:
\begin{eqnarray}
A_h &=&\frac{1}{4} \int d^4x\,  \partial_a h_{bc}\partial_i h_{jk} 
\left[ \eta^{ai}\eta^{bc}\eta^{jk}
 -\eta^{ai}\eta^{bj}\eta^{ck}
+2\eta^{ak} \eta^{bj}\eta^{ci}
 -2\eta^{ak}\eta^{bc}\eta^{ij}\right]\nonumber\\
 &=& \frac{1}{4} \int d^4x\, \left[ \partial_i h^a_a \partial^i h^j_j 
 - \partial_a h_{bc} \partial^a h^{bc} 
 + 2 \partial_a h_{bc} \partial^c h^{ba}
 - 2 \partial_a h^b_b \partial_i h^{ia}
 \right]\nonumber\\
 &=& \frac{1}{4} \int d^4x\, \left[ \frac{1}{2}\partial_i \bar h^a_a \partial^i \bar h^j_j 
 - \partial_a \bar h_{bc} \partial^a \bar h^{bc} 
 + 2 \partial_a \bar h_{bc} \partial^c \bar h^{ba}
 \right]; \qquad \bar h_{ab}\equiv h_{ab}-\frac{1}{2}\eta_{ab}h^i_i
\label{spintwoactapp}
\end{eqnarray}
We shall use the shorter notation:
\begin{equation}
A_h =\frac{1}{4} \int d^4x\,  \partial_a h_{bc}\partial_i h_{jk} M^{abcijk}(\eta^{mn})
\label{spintwoapp}
\end{equation}
where the tensor $M^{abcijk}(\eta^{mn})$  is symmetric in $bc, jk$ and under the triple exchange
$(a,b,c) \leftrightarrow (i,j,k)$ and is given by:
\begin{equation}
M^{abcijk}(\eta^{mn}) = \left[\eta^{ai} \eta^{bc}\eta^{jk}
 -\eta^{ai}\eta^{bj}\eta^{ck}
+2 \eta^{ak}\eta^{bj}\eta^{ci}
 -2\eta^{ak}\eta^{bc}\eta^{ij}\right]_{\rm symm}
 \label{formofmapp}
\end{equation}
where the subscript ``symm'' indicates that the expression inside the square bracket should be
suitably symmetrized in $bc, jk$ and under the exchange $(a,b,c) \leftrightarrow (i,j,k)$.
In the expression for the action, since $M^{abcijk}$ is multiplied by  $\partial_a h_{bc}\partial_i h_{jk}$,
we need not worry about symmetrization and use the expression given inside the square bracket
in Eq.~(\ref{formofmapp}) as it is. 

\subsection {Gauge conditions and true degrees of freedom}

The gauge invariance which was imposed to obtain the action in Eq.(\ref{spintwoactapp}) implies that
 we are dealing with redundant degrees of freedom in $h_{ab}$ and
 --- without additional restrictions --- it does not  carry pure spin-2, in the sense of irreducible representations of 
 Lorentz group. To see this explicitly, consider a $h_{ik}$ of the form:
\begin{equation}
h_{ik}(x)=Q_{ik}(x) + \partial_i A_k(x) + \partial_k A_i(x) + \left( \partial_i \partial_k - \frac{1}{4} \eta_{ik} \partial^2\right)
\alpha(x) + \frac{1}{4}\eta_{ik}\beta(x)
\end{equation} 
 where 
 \begin{equation}
 \partial_i Q^{ik} =0; \qquad \eta_{ik} Q^{ik}=0; \qquad \partial_i A^i =0
 \end{equation} 
  so that $h_{ik}$ (10 components)  is separated into a transverse traceless tensor ($10-5 =5$ components),
  a transverse vector  ($4-1 = 3$ components) and two scalars  $\alpha$ and $\beta$.
  The action in Eq.~(\ref{spintwoactapp}) now becomes 
  \begin{equation}
  A_h = -\frac{1}{2} \int d^4 x\, \left[ \partial_a Q_{bc} \partial^a Q^{bc} - \frac{3}{8} \partial_i \epsilon
  \partial^i \epsilon \right]; \qquad \epsilon\equiv (\beta - \partial^2 \alpha) 
  \end{equation}
   [This is most easily proved in the Fourier space using the third line of Eq.~(\ref{spintwoactapp})].
  This expression shows that: (a) the action is independent of the vector degree of freedom as one would have guessed from the gauge invariance; (b) it does depend on the scalar $\epsilon(x)$ but $\epsilon$
  and $Q^{ab}$ are decoupled from each other; (c) 
  the residual scalar appears with the wrong sign for the kinetic energy term.
  Therefore, isolating the physical degrees of freedom requires imposing the conditions:
 $h^a_a =0$ [to set $\epsilon=0$] and  $\partial_a h^a_b=0$ [to ensure transverse-traceless condition on 
 $Q^{ab}$.]
  If we  impose these gauge conditions  {\it in the action itself}, it becomes:
  \begin{equation}
A_h =\frac{1}{4} \int d^4x\,  \partial_a h_{bc}\partial_i h_{jk} 
\left[ 
 -\eta^{bj}\eta^{ck}\eta^{ai}
+2 \eta^{bj}\eta^{ci}\eta^{ak}
\right]
 =\frac{1}{4} \int d^4x\, \left[ 
 - \partial_a h_{bc} \partial^a h^{bc} 
 + 2 \partial_a h_{bc} \partial^c h^{ba}
 \right]
\label{unimod}
\end{equation}
We mentioned before that the original action in Eq.(\ref{spintwoactapp}) there was an ambiguity with respect to the two terms $c_3$ and $c_4$. 
This non uniqueness disappears in Eq.(\ref{unimod}). There is no possibility of ``swapping"the derivatives
in the first term [since both are $\partial_a$]; in the second term, swapping the derivatives will give a vanishing term
because $\partial_ah^{ba}=0$. We shall see in Appendix B that this action generalizes to an interesting nonlinear theory.   (The five degrees of freedom in $Q^{ab}$ can be further reduced to the two physical degrees of freedom
  of a graviton by using a residual gauge transformation of the form in Eq.(\ref{gtapp}) with $\Box \xi^a=0$; see
  e.g., page 946 of \cite{mtw}; we will not require this result in our discussion). 

\subsection {Gauge invariance and conservation of the source}

The symmetries of the theory are easier to see in the momentum space, which can be done by introducing the Fourier components $f_{ab}(p)$ 
of $h_{ab}(x)$ defined as usual by:
\begin{equation}
h_{ab}(x) \equiv \int \frac{d^4p}{(2\pi)^4} f_{ab}(p) e^{ipx}
\end{equation}
The action becomes 
\begin{equation}
A_h = \frac{1}{4}  \int \frac{d^4p}{(2\pi)^4} f_{bc}  f^*_{jk} \left[p_a p_i M^{abcijk}(\eta^{mn})\right]
\equiv \frac{1}{4}  \int \frac{d^4p}{(2\pi)^4} f_{bc}  f^*_{jk}N^{bcjk}
\label{fourieract}
\end{equation}
with
\begin{eqnarray}
N^{bcjk} &=&\left[ p^2\left(\eta^{bc} \eta^{jk} - \eta^{bj}\eta^{ck}\right) + 2 p^k \left( p^c \eta^{bj} - p^j \eta^{bc} \right) \right]_{\rm symm}  \nonumber\\
&=& \left( p^2 \eta^{bc} \eta^{jk} - p^k p^j \eta^{bc} - p^c p^b \eta^{jk} \right)  
 - \frac{p^2}{2}   \left(\eta^{bj} \eta^{ck} + \eta^{cj}\eta^{bk}\right) 
  + \frac{1}{2} \left( p^k p^c \eta^{bj} + p^j p^c \eta^{bk} + p^k p^b \eta^{cj} + p^j p^b \eta^{ck}\right)
  \label{glory}
\end{eqnarray}
where we have exhibited the  symmetrized expression in full glory for once. 
[$N^{bcjk}$ is symmetric in $bc,jk$ and under the pair exchange $(b,c) \leftrightarrow (j,k)$.]
The gauge transformation of the spin-2 field, given  by Eq.(\ref{gtapp}) is equivalent to
 $f_{ab} \to f_{ab} + p_a \xi_b + p_b \xi_a$ in the Fourier space.  Using this in Eq.~(\ref{fourieract})
we get 
\begin{equation}
A_h \to  \frac{1}{4}  \int \frac{d^4p}{(2\pi)^4}( f_{bc} + 2 p_b\xi_c) ( f^*_{jk} + 2 p_j \xi^*_k)  N^{bcjk}
\end{equation}
Straight forward computation now shows that
$p_b N^{bcjk} =0 = p_j N^{bcjk} $ making
 $A_h$ is invariant under the gauge transformations. This is, of course, built-in by the choice of the coefficients $c_i$ in the original action. This condition translates, in coordinate space, to the {\it identity}
 $M^{abcijk} \partial_b\partial_a\partial_i h_{jk}=0$.

\section*{Appendix B: Comment on the uniqueness}

We briefly comment on two issues in this Appendix, postponing their detailed discussion to a future publication.

First, it may be noted that the action for spin-2 graviton had an ambiguity in the form a four divergence
term $A_{\rm div}$ in Eq. (\ref{divamb}). While the action in Eq. (\ref{spintwo}) correctly leads to full Einstein theory under
the substitution $d^4x\to d^4x\sqrt{-g}, \eta_{ab}\to g_{ab}=\eta_{ab}+\lambda h_{ab}$, the four divergence term
in Eq.(\ref{divamb}) does not map to a four divergence term under these substitutions:
\begin{equation}
A_{\rm  div}=
\frac{c_3}{4} \int d^4x\,  \partial_a h_{bc}\partial_i h_{jk} 
\left[ 
 \eta^{ab}\eta^{ck}\eta^{ij}
-\eta^{ak} \eta^{bj}\eta^{ci}
 \right]
\to
\frac{c_3}{4} \int d^4x\sqrt{-g}\,  \partial_a g_{bc}\partial_i g_{jk} 
\left[ 
 g^{ab}g^{ck}g^{ij}
-g^{ak} g^{bj}g^{ci}
 \right]
\end{equation}
which cannot be expressed as a four divergence. So, we obtain the correct Einstein's theory only if we start with
the correct set of terms in the linear spin-2 case.  In general, consider  two different actions $A_I^{lin}$ and $A_{II}^{lin}$ in the linear theory, which differ by a four divergence. The substitutions 
$d^4x\to d^4x\sqrt{-g}, \eta_{ab}\to g_{ab}=\eta_{ab}+\lambda h_{ab}$
 can be used with either $A_I^{lin}$ or $A_{II}^{lin}$
  and will  lead to  nonlinear theories with actions $A_I^{nl}$ or $A_{II}^{nl}$.
 But $A_I^{nl}$ and $A_{II}^{nl}$
 cannot be related by a four-divergence, in general. 

Second,  let us consider the spin-2 theory obtained by imposing the gauge conditions $h^a_a =0, \partial_a h^a_b=0$ in the action itself. The linear theory action was given by 
Eq. (\ref{unimod}) and, as was pointed out before,
there is no ambiguity as regards four divergences in this action.  Our ``rule" now leads to
 the nonlinear action:
 \begin{equation}
A_\infty 
= \frac{1}{4\lambda^2} \int d^4x\, \sqrt{-g}\,  \partial_a g_{bc}\partial_i g_{jk} 
\left[ 
 -g^{bj}g^{ck}g^{ai}
+2 g^{bj}g^{ci}g^{ak}
 \right]
\label{finalunimod}
\end{equation}  
To get the field equations from this action, we only need to note that it is the same as the
standard $\Gamma^2$ action in Eq~(\ref{final}) with the extra condition $g_{ab}dg^{ab}=0$. Hence the field equations resulting from varying $g_{ab}$ in Eq.~(\ref{finalunimod}) will be the same as those obtained from
 varying $g_{ab}$ in the action in Eq~(\ref{final}), 
 keeping $\sqrt{- g}=$ constant. It is well known that, we then get the equations:
 \begin{equation}
      R_{ab} - \frac{1}{4} g_{ab} R=8\pi(T_{ab} - \frac{1}{4} g_{ab} T)
      \label{unimodeqn}
      \end{equation}
      in which both sides are trace free. Bianchi identity can now be used to show that
      $\partial_a(R+8\pi T)=0$, requiring $(R+8\pi T)=$ constant. Thus  cosmological constant arises
      as an (undetermined) integration constant in such models \cite{tppr}, and could be interpreted as a
      Lagrange multiplier that maintains the condition $\sqrt{-g}=$ constant.  Clearly, the gauge conditions
       translate to $\sqrt{-g}=$ constant in the full theory eliminating the scalar degree of freedom.

It should be stressed that, while this theory is mathematically the same as Einstein's gravity with a cosmological constant, it is conceptually quite different from the usual approach to cosmological constant. 
What is probably more interesting,
this theory takes gravity one step closer to gauge theories in the following sense: It has been known for a long time that
the Christoffel symbols in gravity $\Gamma_{ab}^c$ can be thought of as the elements of the matrix $(\Gamma_a)^c_b$
in exact analogy with the matrix representation of the gauge field $(A_i)^j_k$. The Riemann-Christoffel tensor can then be given the matrix representation
\begin{equation}
R_{ab}=\partial_a \Gamma_b-\partial_b \Gamma_a +\Gamma_a\Gamma_b-\Gamma_a\Gamma_b
\end{equation}
with two matrix indices suppressed etc. Everyone who followed this route soon realized that --- unfortunately ---
we need to contract on a matrix index with spacetime index to get the Einstein action etc. In the above approach,
if we take $\sqrt{-g}=1$ then the quadratic action can be expressed in the form
\begin{equation}
A_{\rm quad}=\frac{1}{4\lambda^2}\int d^4x\,  g^{ab}Tr[\Gamma_a \Gamma_b];\qquad \sqrt{-g}=1
\end{equation}
We hope to discuss the gauge theory connection arising from this approach in a future publication.

\end{document}